\documentclass[a4paper,english,aps,prb,twocolumn]{revtex4}
\usepackage[T1]{fontenc}
\usepackage[latin1]{inputenc}
\usepackage{subfigure}
\usepackage{graphicx}

\makeatletter

\providecommand{\tabularnewline}{\\}

\usepackage{babel}
\makeatother
\begin{document}

\preprint{This line only printed with preprint option}

\title{Electronic Structure and Electron-Phonon Interaction in Hexagonal
Yttrium }

\author{Prabhakar P. Singh}

\email{ppsingh@phy.iitb.ac.in}

\affiliation{Department of Physics, Indian Institute of Technology, Powai, Mumbai-
400076, India }

\begin{abstract}
To understand the pressure-induced changes in the electronic structure
and the electron-phonon interaction in yttrium, we have studied hexagonal
close-packed \emph{}(hcp) yttrium, stable at ambient pressure and
double hexagonal close-packed \emph{}(dhcp) yttrium, stable up to
around $44$ GPa, using density-functional-based methods. Our results
show that as one goes from hcp yttrium to dhcp yttrium, there is (i)
a substantial charge-transfer from $s\rightarrow d$ with extensive
modifications of the $d$-band and a sizable reduction in the density
of states at the Fermi energy, (ii) a substantial stiffening of phonon
modes with the electron-phonon coupling covering the entire frequency
range, (iii) an increase in the electron-phonon coupling constant
$\lambda$ from $0.55$ to $1.24$, leading to a change in the superconducting
transition temperature $T_{c}$ from $0.3$ K to $15.3$ K for $\mu^{*}=0.2$. 
\end{abstract}
\maketitle

\section{introduction}

The change in the superconducting transition temperature $T_{c}$
in yttrium from $6$ mK at ambient pressure \cite{probst} to around
$20$ K at $115$ GPa \cite{hamlin}, concurrent with the change in
its crystal structure \cite{gro_dhcp,vohra}, underscores the intricate
relationship between pressure-induced $s\rightarrow d$ charge transfer,
dynamical response of the lattice and its coupling to Fermi electrons.
The experiment on yttrium follows the recent observation of superconductivity
in compressed lithium at a $T_{c}$ of $20$ K \cite{shimuzu,stru},
bringing into focus the pressure-induced changes in the electron-phonon
interactions in elemental metals, especially the alkali and the early-transition
metals \cite{deepa,profeta,sanna}. 

Experimentally, yttrium follows the equilibrium structure sequence
hcp$\rightarrow$Sm-type$\rightarrow$dhcp$\rightarrow$trigonal as
the pressure is increased from ambient to $50$ GPa at room temperature
\cite{gro_dhcp}. It is expected that yttrium will stabilize in a
body-centered cubic structure at pressures above $280$ GPa \cite{melsen}.
On the other hand, The superconducting transition temperature $T_{c}$
changes from $6$ mK at ambient pressure \cite{probst} to around
$1.3$ K at $11$ GPa \cite{wittig}, and then to $17$ K at $89$
GPa \cite{hamlin}. A further increase in pressure to $115$ GPa results
in a $T_{c}=20$ K. Surprisingly, it seems possible that for pressures
greater than $115$ GPa, yttrium could superconduct with a $T_{c}$
higher than that of $20$ K.

In an effort to understand the pressure-induced changes in the superconducting
properties of yttrium \emph{vis-a-vis} the changes in its stable crystal
structure, we have studied, using density-functional based methods,
(i) the electronic structure, (ii) the phonon density of states, (iii)
the electron-phonon interaction, and (iv) the solutions of the isotropic
Eliashberg gap equation for hcp and dhcp yttrium at ambient pressure
and at around $50$ GPa. 

We have calculated the electronic structure of hcp and dhcp yttrium
using full-potential, linear muffin-tin orbital (LMTO) method \cite{savrasov1,savrasov2}
as well as the plane-wave pseudopotential method using Quantum-ESPRESSO
package \cite{pwscf}. Using the linear response code based on the
LMTO method,  we have calculated the phonon DOS, $F(\omega)$, and
the Eliashberg function, $\alpha^{2}F(\omega)$, for hcp and dhcp
yttrium. Subsequently, we have numerically solved the isotropic Eliashberg
gap equation \cite{allen1,allen2,private1} for a range of $\mu^{*}$
to obtain the corresponding superconducting transition temperature
$T_{c}.$ After the completion of the present work, we have come across
the work of Yin \emph{et al}. \cite{yin}, which uses the \emph{hypothetical}
fcc structure to describe the effects of pressure in yttrium. 

Based on our calculations, described below, we find that as one goes
from hcp yttrium to dhcp yttrium, there is (i) a substantial charge-transfer
from $s\rightarrow d$ with extensive modifications of the \emph{d}-band,
(ii) a stiffening of phonon modes with the electron-phonon coupling
over the entire frequency range, (iii) an increase in electron-phonon
coupling constant $\lambda$ from $0.55$ to $1.24$, leading to a
change in the superconducting transition temperature $T_{c}$ from
$0.3$ K to $15.3$ K for $\mu^{*}=0.2,$ consistent with experiment.

\section{Theoretical Approach and computational details}

The density-functional theory provides a reliable framework for implementing
from first-principles the Migdal-Eliashberg approach for calculating
the superconducting properties of metals. Such an approach has been
implemented using the full-potential LMTO method \cite{savrasov1,savrasov2}. 

In general, the phonon density of states is given by \[
F(\omega)=\sum_{\mathbf{q}\nu}\delta(\omega-\omega_{\mathbf{q}\nu}),\]
where $\omega_{\mathbf{q}\nu}$ is the $q\nu$-th phonon mode associated
with the atomic displacements. Using the electron-phonon matrix elements
$g_{\mathbf{k}+\mathbf{q}j',\mathbf{k}j}^{\mathbf{q}\nu}$, which
can be interpreted as the scattering of electron in state $|\mathbf{k}j>$
to state $|\mathbf{k+q}j'>$ due to perturbation arising out of the
phonon mode $\omega_{\mathbf{q}\nu}$, we can calculate the phonon
line width $\gamma_{\mathbf{q}\nu}$

\[
\gamma_{\mathbf{q}\nu}=2\pi\omega_{\mathbf{q}\nu}\sum_{\mathbf{k}jj'}|g_{\mathbf{k}+\mathbf{q}j',\mathbf{k}j}^{\mathbf{q}\nu}|^{2}\delta(\epsilon_{\mathbf{k}j}-\epsilon_{F})\delta(\epsilon_{\mathbf{k+q}j'}-\epsilon_{F}).\]
where $N(\varepsilon_{F})$ is the electronic density of states at
the Fermi energy. Now, we can combine the electronic density of states,
phonon spectrum and the electron-phonon matrix elements to obtain
the Eliashberg function $\alpha^{2}F(\omega)$ defined as 

\[
\alpha^{2}F(\omega)=\frac{1}{2\pi N(\epsilon_{F})}\sum_{\mathbf{q}\nu}\frac{\gamma_{\mathbf{q}\nu}}{\omega_{\mathbf{q}\nu}}\delta(\omega-\omega_{\mathbf{q}\nu})\]
Finally, the Eliashberg functions can be used to solve the isotropic
gap equation \cite{allen1,allen2,private1}\[
\Delta(i\omega_{n})=\sum_{n'}^{|\omega_{n'}|<\omega_{c}}f_{n'}S(n,n')\Delta(i\omega_{n'})\]
to obtain the superconducting properties such as the superconducting
transition temperature $T_{c}.$ The function $S(n,n')$ used in the
gap equation is defined by \[
S(n,n')\equiv\lambda(n-n')-\mu^{*}-\delta_{nn'}\sum_{n''}s_{n}s_{n''}\lambda(n-n'')\]
and $f_{n}=1/|2n+1|$ with $s_{n}$ representing the sign of $\omega_{n}.$
The electron-phonon coupling $\lambda(\nu)$ is given by \[
\lambda(\nu)=\int_{0}^{\infty}d\omega\alpha^{2}F(\omega)\frac{2\omega}{\omega_{\nu}^{2}+\omega^{2}}.\]

Before describing our results, we provide some of the computational
details of the present calculations.

The self-consistent electronic structures of hcp and dhcp yttrium
in $P6_{3}/mmc$ crystal structure, having two and four yttrium atoms
in the respective primitive cells, were calculated using the pwscf
code of the Quantum-ESPRESSO package and the full-potential, LMTO
method. The results obtained with the pwscf code are used to study
the equation of state of yttrium. All other properties of yttrium,
described below, have been obtained with the full-potential, LMTO
method and linear response code based on the LMTO method. 

The plane-wave pseudopotential method with Vanderbilt's ultrasoft
pseudopotentials were used to calculate the electronic structure of
hcp and dhcp yttrium with 259 $\mathbf{k}$-points in the irreducible
Brillouin zones. The ultrasoft pseudopotential used also included
the 4s and 4p states with nonlinear core correction. The kinetic energy
cutoff for the wavefunction was taken to be $35$ Ry, while for the
charge density and the potential the kinetic energy cutoff were $200$
Ry. For exchange-correlation potential we used the local-density approximation
as given by Perdew \emph{et al} \cite{perdew1}. These calculations
were used to study the equation of state of yttrium.

The charge self-consistent, full-potential, linear muffin-tin orbital
calculations for hcp and dhcp yttrium were carried out with the local-density
approximation for exchange-correlation of Perdew \emph{et al} \cite{perdew1}
with $2\kappa$-energy panels and 610 and 549 $\mathbf{k}$-points
respectively, in the corresponding irreducible wedges of the hexagonal
Brillouin zones. The 4\emph{s} state of yttrium was treated as a semicore
state, while the 4\emph{p} state was treated as a valence state. The
basis set used consisted of $s,$ $p,$ and $d$ orbitals at the yttrium
site, and the potential and the wave function were expanded up to
$l_{max}=6$. The muffin-tin radii for yttrium at zero pressure and
at $50$ GPa were taken to be $3.3$ and $2.8$ atomic units, respectively.
These calculations were used to study the \emph{s}-, \emph{p}- and
\emph{d}-resolved densities of states, the band-structure along the
high symmetry directions in the hexagonal Brillouin zones and the
Fermi surfaces of hcp and dhcp yttrium. The Fermi surfaces of hcp
and dhcp yttrium were constructed using XCrySDen program \cite{xcrys}
with eigenvalues calculated on a $36\times36\times24$ and $36\times36\times12$
grid in the reciprocal space respectively.

The phonon density of states and the Eliashberg function of hcp and
dhcp yttrium were calculated using the  linear response code based
on the full-potential, LMTO method. The dynamical matrices and the
electron-phonon matrix elements of hcp and dhcp yttrium were calculated
on a $6\times6\times4$ and $6\times6\times2$ grids respectively,
resulting in $21$ and $14$ irreducible $\mathbf{q}$-points. The
Brillouin zone integrations during linear response calculations for
hcp and dhcp yttrium were carried out using a $12\times12\times8$
and $12\times12\times4$ grids of $\mathbf{k}$-points. The Fermi
surface sampling for the evaluation of the electron-phonon matrix
elements for the two structures were done using $36\times36\times24$
and $36\times36\times12$ grids.

\section{results and discussion}

In this section we describe the results of our calculations of the
electronic structure, the linear response and the solutions of the
isotropic Eliashberg gap equation for hcp and dhcp yttrium. Our results
are described in terms of (i) equation of state, (ii) electronic structure,
(iii) electron-phonon interaction, and (iv) superconducting transition
temperature obtained from the solutions of the isotropic Eliashberg
gap equation. 

\begin{figure}
\begin{centering}\includegraphics[clip,width=7.4cm,height=7.4cm]{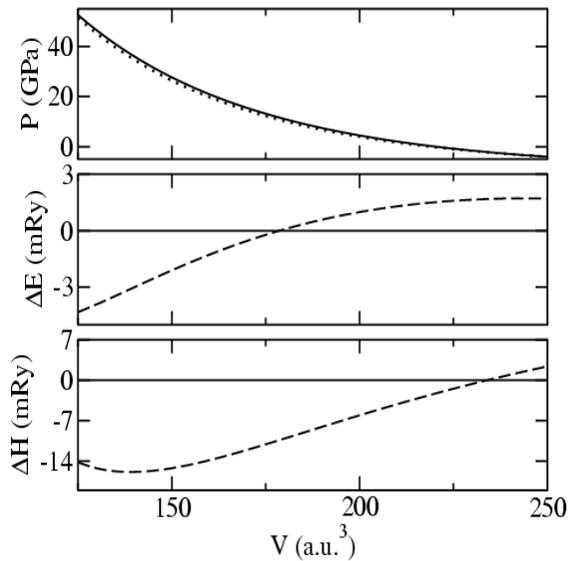}\par\end{centering}

\caption{The equation of state (top panel) of hcp (solid line) and dhcp (dotted
line) yttrium, the difference in energy $\Delta E$ (middle panel,
dashed line) and enthalpy $\Delta H$ (bottom panel, dashed line)
of dhcp yttrium with respect to hcp yttrium as a function of volume
per atom. }
\end{figure}

\subsection{Equation of State}

The calculated equations of state of yttrium in hcp and dhcp phases
are shown in Fig. 1. These equations of state are obtained by fitting
the calculated total energy versus volume points to the third-order
Birch-Murnaghan equation of state \cite{birch}. The value of $c/a$
was optimized by iteratively minimizing the total energy as a function
of volume. The lattice constants $a$ and $c$ at zero and $50$ GPa
for both hcp and dhcp phases, as obtained through the above procedure,
are given in Table I. However, the lattice constant $a$ for the high-pressure
dhcp phase of yttrium corresponds to the compressed volume equal to
$0.56v_{0}$, with the ideal $c/a=3.25$ and where $v_{0}$ is the
experimentally determined equilibrium volume of the hcp phase. The
lattice constants as given in Table I have been used in all the calculations
reported below. We also note that in our discussion, we will emphasize
the stable structures, viz. hcp at zero pressure and dhcp at $50$
GPa pressure, more than the unstable structures at these pressures.

\begin{table}

\caption{The lattice constants of hcp and dhcp yttrium used in the calculations.
The experimental values \cite{gro_dhcp} are given in the parentheses.}

\begin{centering}\begin{tabular}{|c|c|c|c|}
\hline 
structure&
pressure (GPa)&
$a$ (a.u.)&
$c/a$\tabularnewline
\hline
\hline 
hcp&
0&
6.775 (6.895)&
1.56 (1.571)\tabularnewline
\hline 
hcp&
50&
5.604&
1.65\tabularnewline
\hline 
dhcp&
0&
6.811&
3.20\tabularnewline
\hline 
dhcp&
50&
5.619&
3.25\tabularnewline
\hline
\end{tabular}\par\end{centering}
\end{table}

The calculated lattice constants for hcp yttrium at zero pressure,
as given in Table I, are somewhat smaller than the experimental values,
as expected in local-density approximation. We find the bulk modulus
of hcp yttrium at equilibrium to be $39.4$ GPa, which compares well
with the experimental value of $41$ GPa \cite{gro_dhcp}. In Fig.
1, we also show the energy and enthalpy of dhcp yttrium with respect
to hcp yttrium as a function of volume. At zero pressure the dhcp
phase is approximately $3$ mRy higher in energy than the hcp phase.
However, with increase in pressure both energy and enthalpy of the
dhcp phase becomes lower than the hcp phase, in agreement with experiment.

\begin{figure}
\begin{centering}\subfigure[]{\includegraphics[scale=0.33,angle=270]{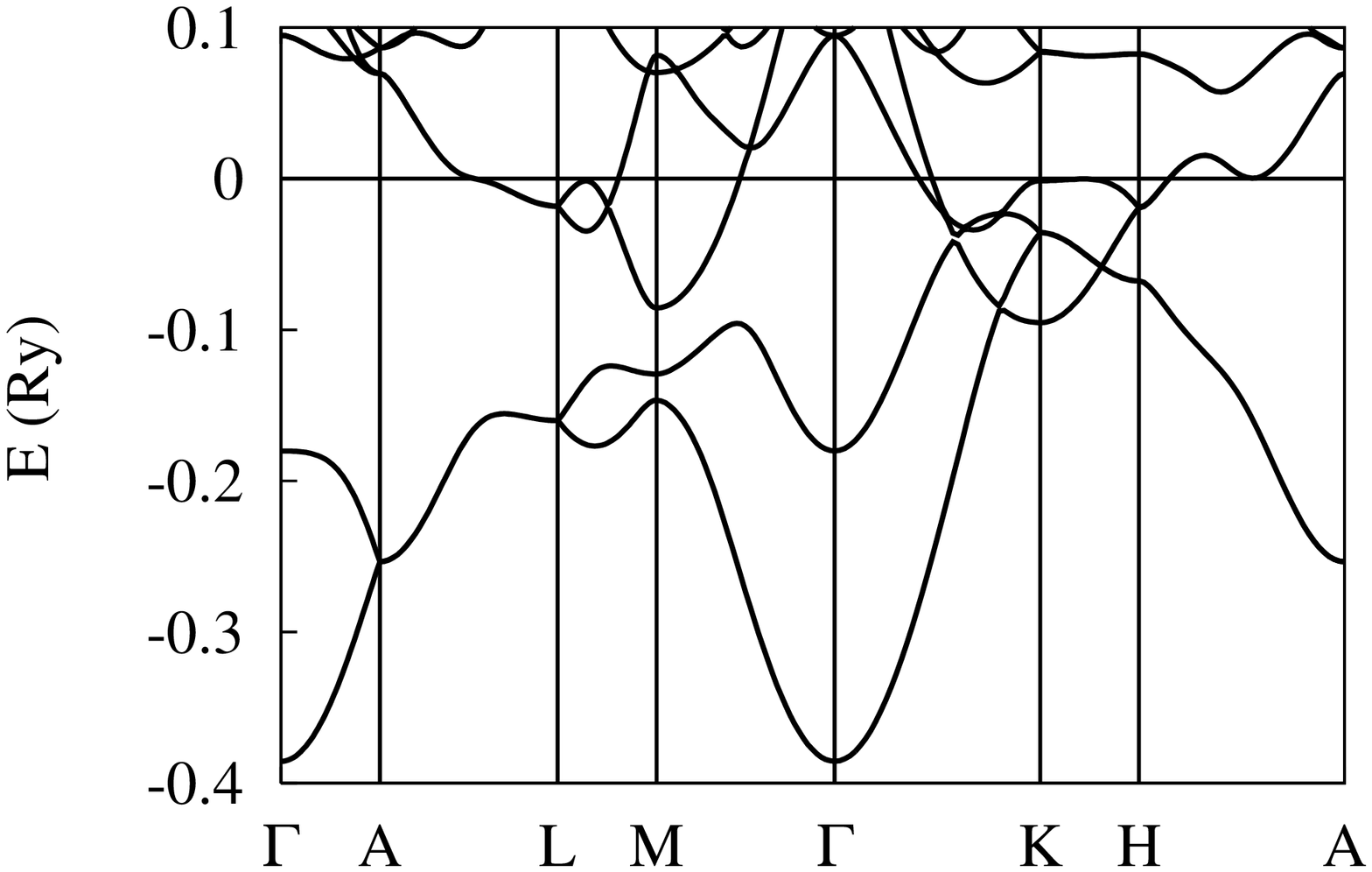}}\par\end{centering}

\begin{centering}\subfigure[]{\includegraphics[scale=0.33,angle=270]{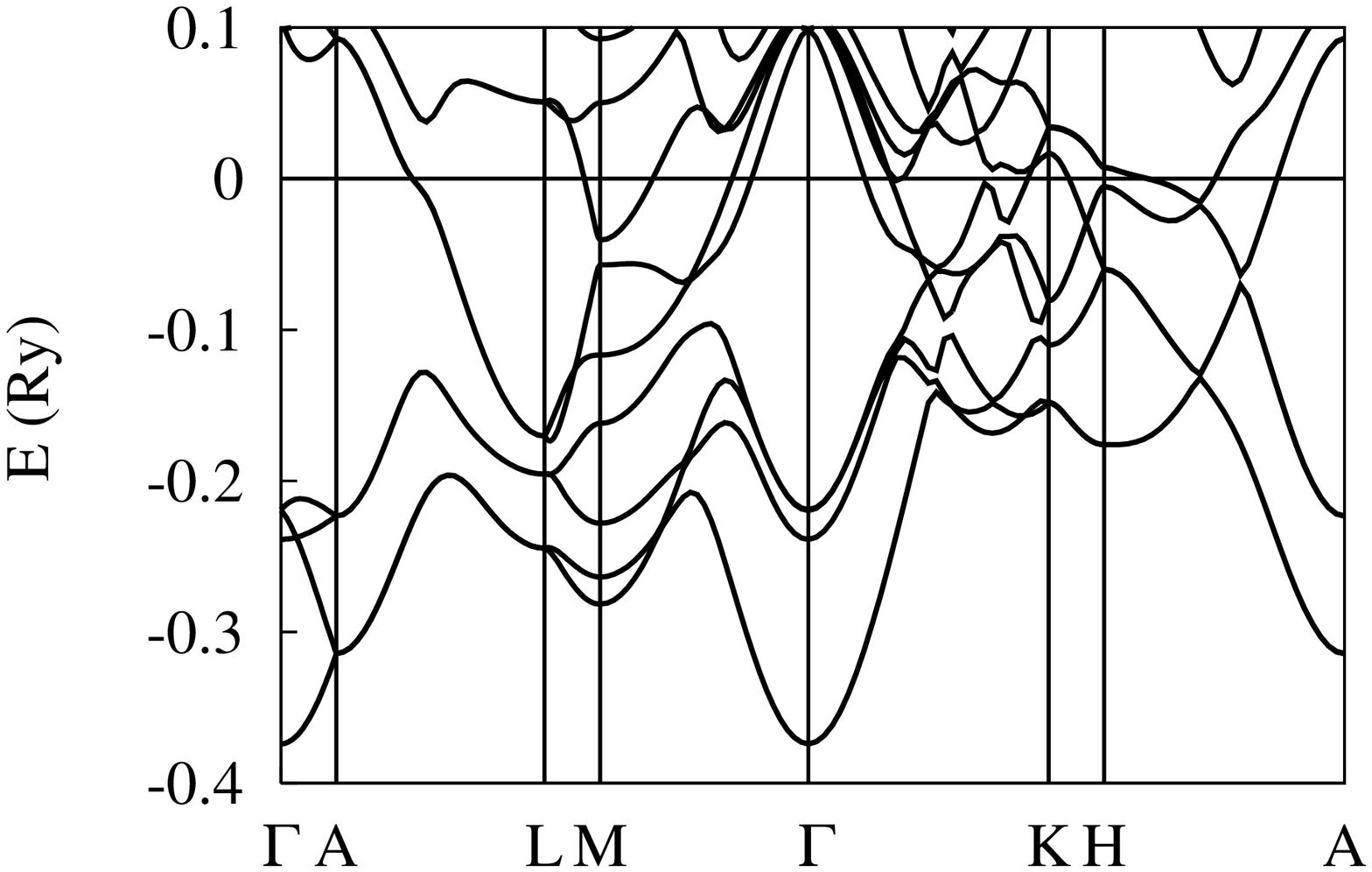}}\par\end{centering}

\begin{centering}\subfigure[]{\includegraphics[scale=0.33,angle=270]{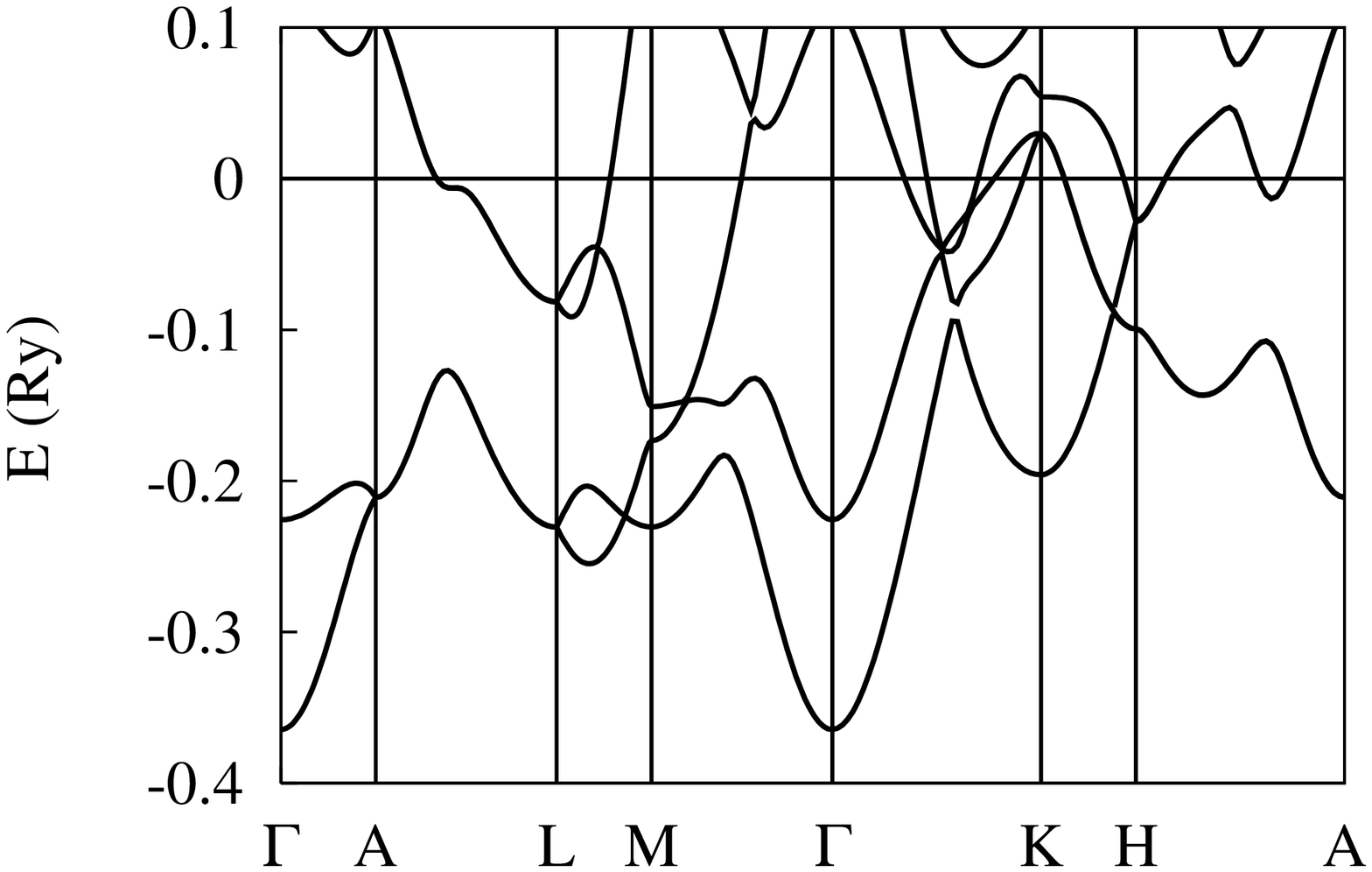}}\par\end{centering}

\begin{centering}\includegraphics[scale=0.33,angle=270]{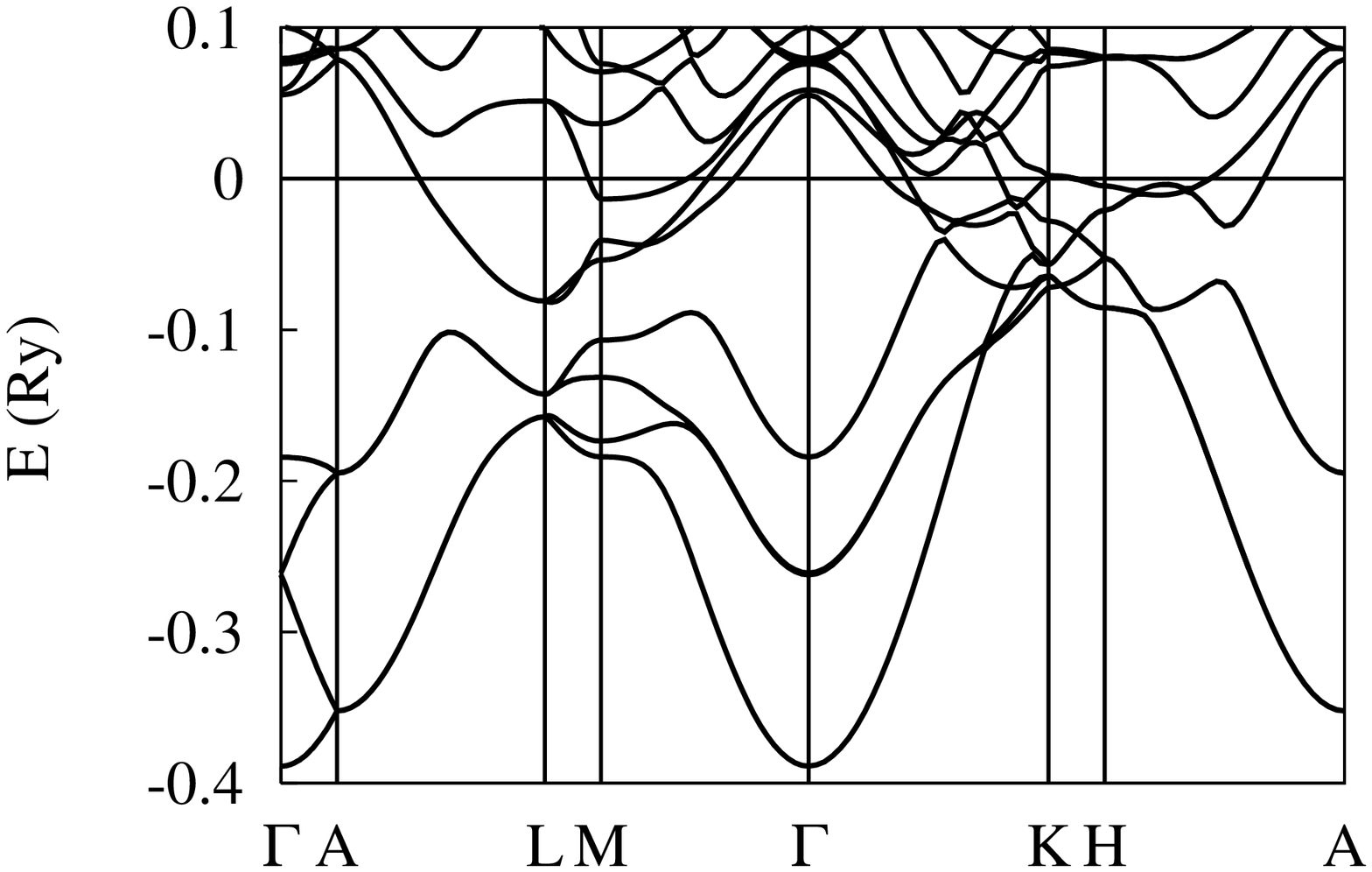}\par\end{centering}

\caption{The band-structure along high symmetry directions in the hexagonal
Brillouin zones of (a) hcp yttrium at ambient pressure, (b) dhcp yttrium
at $50$ GPa, (c) hcp yttrium at $50$ GPa and (d) dhcp yttrium at
ambient pressure. The horizontal line, passing through the energy
zero, indicates the Fermi energy. }
\end{figure}

\subsection{Electronic Structure}

We show in Fig. 2, the calculated band-structure along high symmetry
directions in the hexagonal Brillouin zones of hcp and dhcp yttrium
at both zero and at $50$ GPa pressure. The band corresponding to
4\emph{p} state is much lower in energy, and it is not shown in the
figure. We find that in the case of hcp yttrium at zero pressure only
two bands, the 3rd and the 4th bands in Fig. 2 (a), cross the Fermi
energy. Based on the band-structure, we expect the states close to
the zone boundary such as M, K, H, and L symmetry points to contribute
to the electron-phonon coupling. For dhcp yttrium with 4-atom basis,
we find that there are four bands crossing the Fermi energy at $50$
GPa, as shown in Fig. 2 (b). There are states around symmetry points
M, K and H that are close to Fermi energy, therefore we expect these
states to play an important role in electron-phonon coupling in compressed
dhcp yttrium. Due to charge transfer from $s\rightarrow d$ the bottom
of the \emph{s}-band moves slightly up in dhcp yttrium as can be seen
by comparing Figs. 2 (a) and 2 (b).

In Figs. 2 (c)-(d), we show the band-structure of hcp phase at high
pressure and dhcp phase at ambient pressure, respectively. We find
that the increase in pressure has changed the band-structure of the
hcp phase, especially along $\Gamma$-A and around symmetry points
K and H, in such a way that it looks similar to the high pressure
dhcp phase. In particular, the flat portion along K-H in the hcp phase,
as shown in Fig. 2 (a), has been completely modified at high pressure,
as shown in Fig. 2 (c). 

\begin{figure}
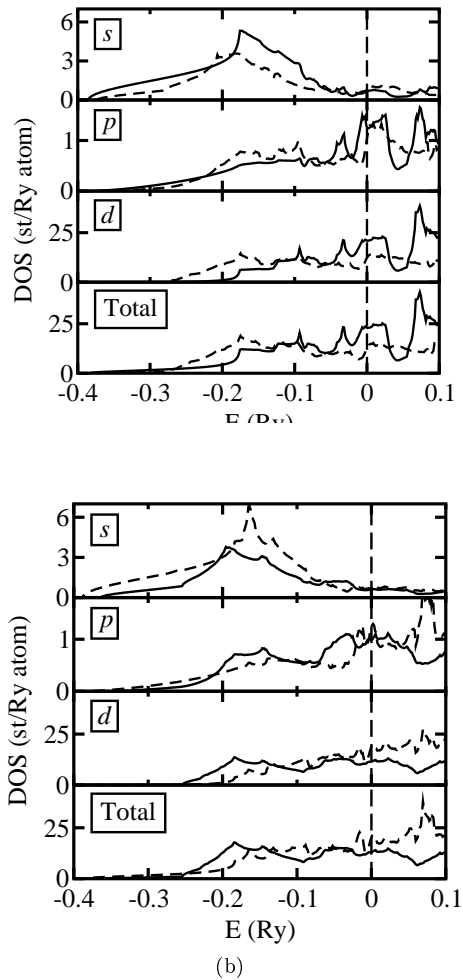

\begin{centering}\subfigure[]{\includegraphics[clip,scale=0.33]{fig3a}}\par\end{centering}

\begin{centering}\subfigure[]{\includegraphics[scale=0.33]{fig3b}}\par\end{centering}

\caption{The \emph{s}, \emph{p}, \emph{d} and the total densities of states
of (a) hcp yttrium at ambient pressure (solid line) and dhcp yttrium
at $50$ GPa (dashed line), and (b) hcp yttrium at $50$ GPa (solid
line) and dhcp yttrium at ambient pressure (dashed line). The vertical,
dashed line indicates the Fermi energy. }
\end{figure}

A more comprehensive picture of $s\rightarrow d$ charge transfer
can be obtained by comparing the densities of states of hcp and dhcp
yttrium. In Fig. 3, we show the  total and the \emph{l}-resolved densities
of states of hcp and dhcp yttrium. By comparing the total densities
of states in the two structures, we find that many more states are
created between $-0.26$ Ry and $-0.16$ Ry in the compressed dhcp
yttrium to accommodate the pushed out \emph{s}-electrons as well as
the \emph{d}-electrons pushed further inside from the flattened portion
of the \emph{d}-band near the Fermi energy. We also find that in the
energy interval $-0.2$ Ry to $-0.05$ Ry, some \emph{p}-states are
also created in the compressed dhcp yttrium. 

The density of states around the Fermi energy plays an important role
in determining the electron-phonon interaction in metals. In the present
case, we find that the total density of states of $22.8$ st/Ry atom
at the Fermi energy in hcp yttrium reduces to $13.4$ st/Ry atom in
dhcp yttrium. The reduction in \emph{s} and \emph{d} states are from
$0.71$ and $20.9$ st/Ry atom in hcp yttrium to $0.6$ and $11.7$
st/Ry atom in dhcp yttrium, respectively. Usually, an increase in
the density of states at the Fermi energy strengthens the electron-phonon
coupling. However, in spite of a decrease in the density of states
at the Fermi energy as yttrium is compressed, there is an increase
in the strength of the electron-phonon coupling. 

The changes in the densities of states of the unstable phases of yttrium,
namely hcp at high pressure and dhcp at ambient pressure, are shown
in Fig. 3 (b). The pressure-induced changes in going from ambient
pressure dhcp phase to high pressure hcp phase are qualitatively similar
to that of hcp to dhcp as described above. For example, the total
density of states of $21.1$ st/Ry atom at the Fermi energy in dhcp
yttrium, at ambient pressure, reduces to $13.1$ st/Ry atom in hcp
yttrium at high pressure. The changes in \emph{s} and \emph{d} states
are from $0.58$ and $19.3$ st/Ry atom in dhcp yttrium to $0.6$
and $11.5$ st/Ry atom in hcp yttrium at high pressure, respectively.

\begin{figure}
\begin{centering}\subfigure[]{\includegraphics[clip,scale=0.25]{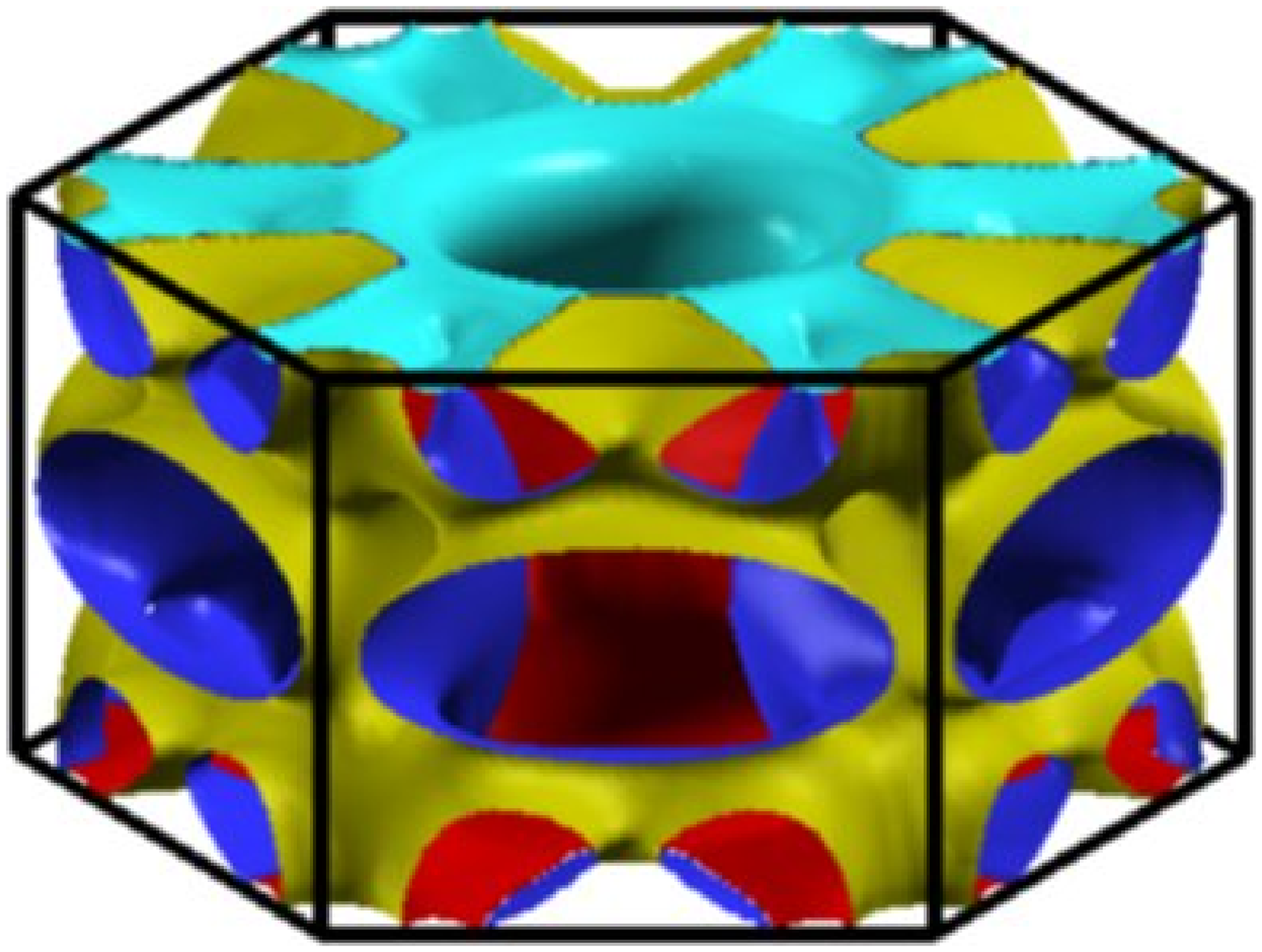}}~
\subfigure[]{\includegraphics[clip,scale=0.25]{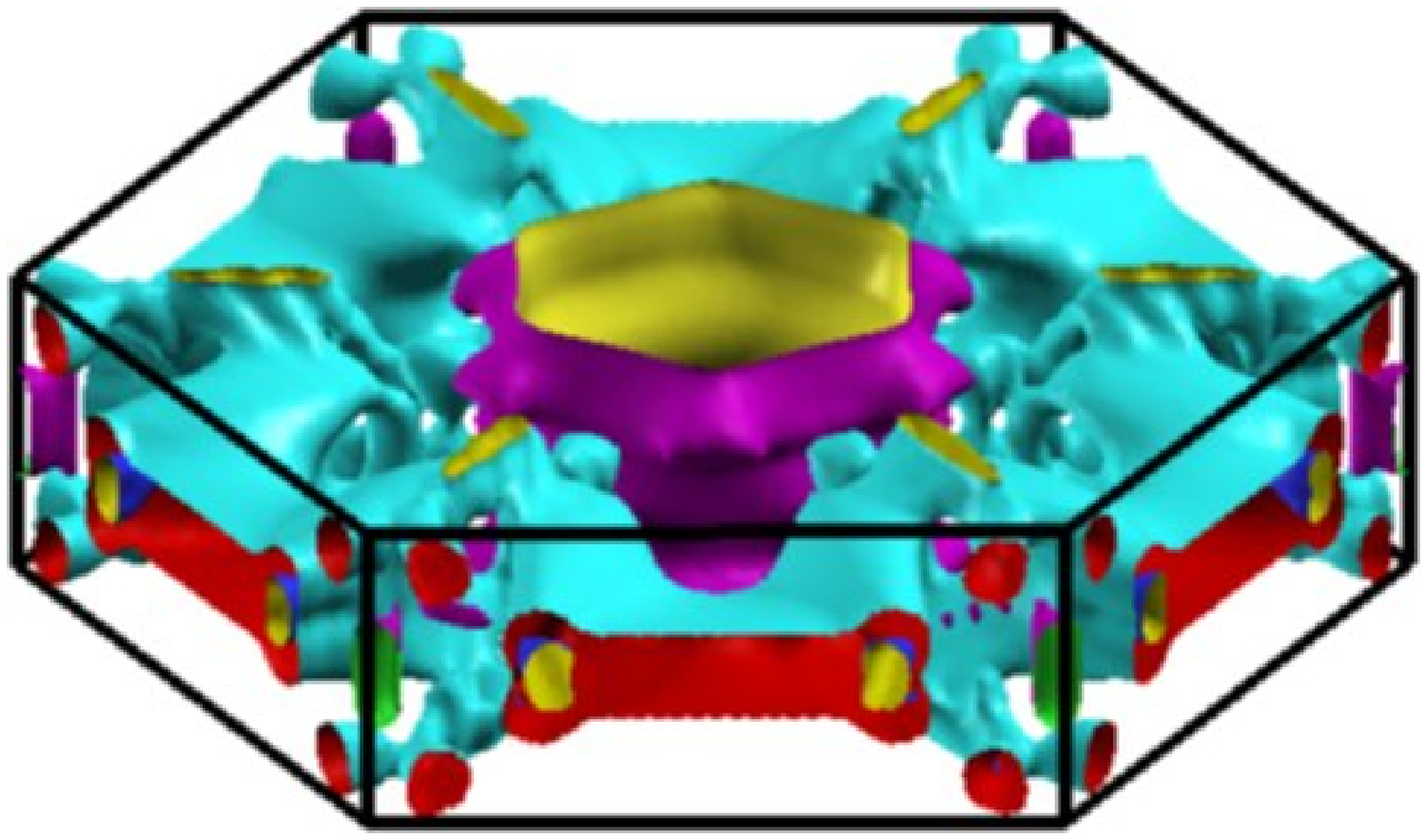}}\par\end{centering}

\begin{centering}\subfigure[]{\includegraphics[clip,scale=0.25]{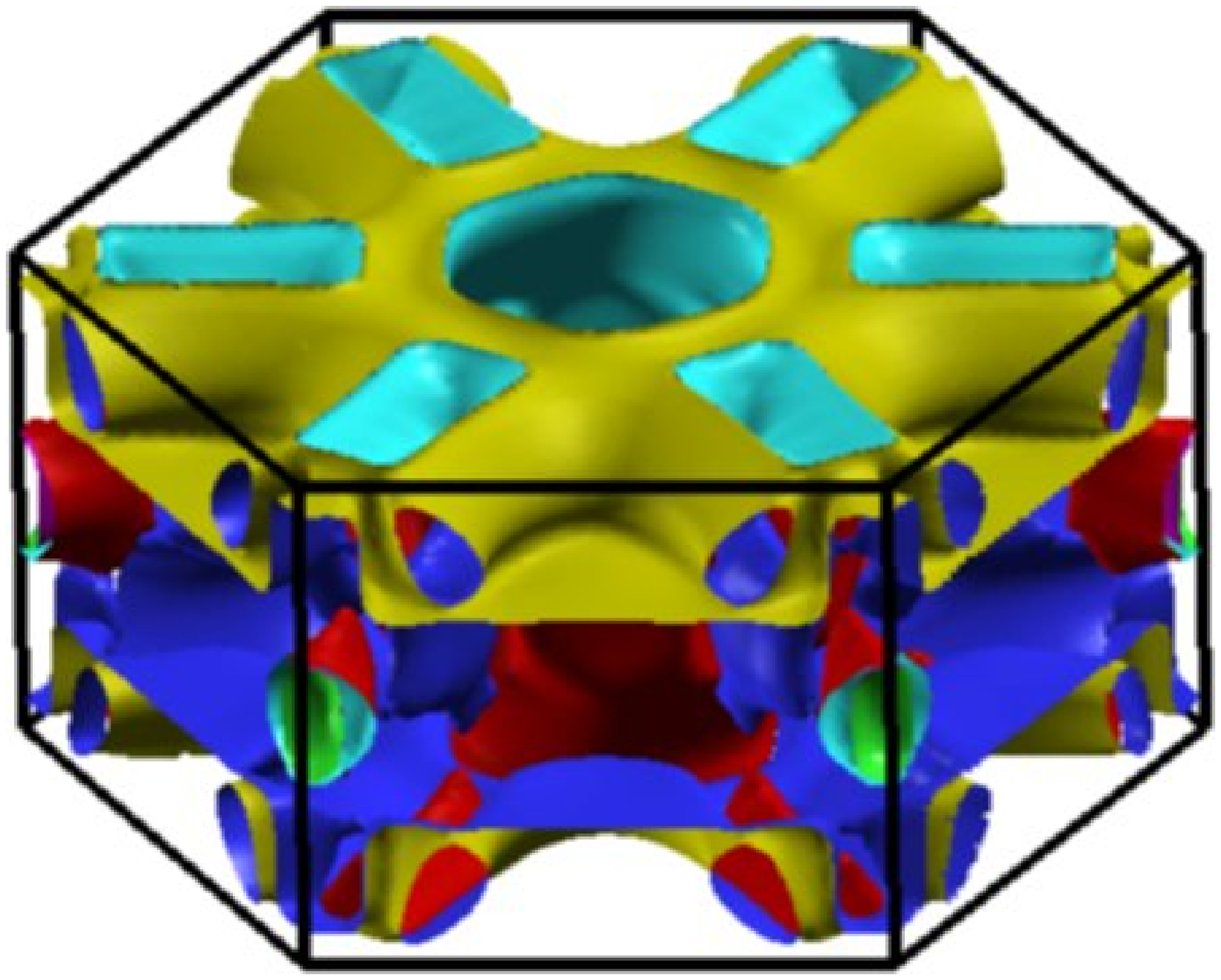}}~\subfigure[]{\includegraphics[clip,scale=0.25]{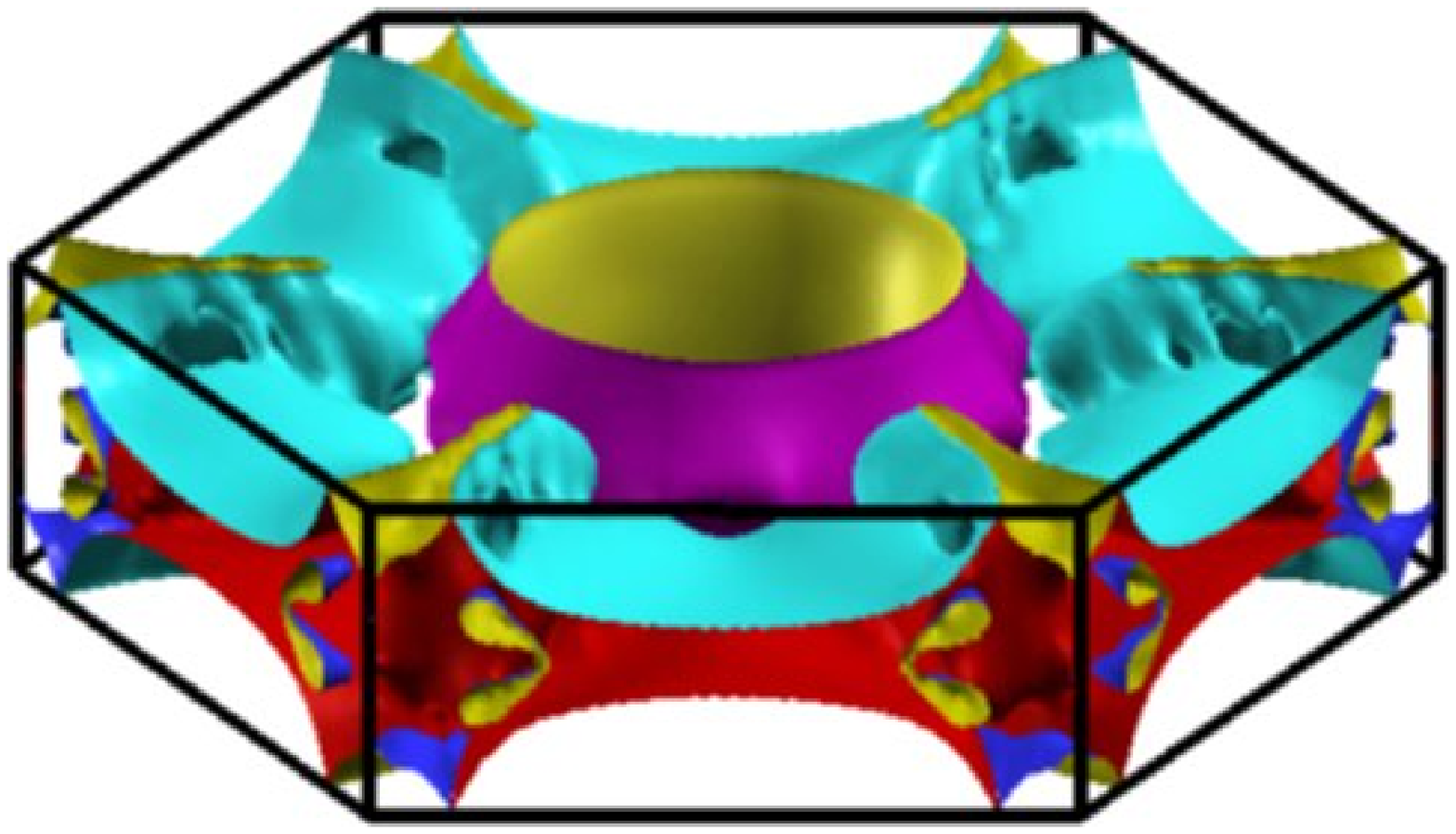}}\par\end{centering}

\caption{The Fermi surface of (a) hcp yttrium at ambient pressure, (b) dhcp
yttrium at $50$ GPa, (c) hcp yttrium at $50$ GPa, and (d) dhcp yttrium
at ambient pressure. }
\end{figure}

The topology of the Fermi surface is useful in understanding many
of the reciprocal-space-based properties of the solids. In particular,
flat portions of the Fermi surface can give rise to peaks in the response
function, leading to a substantial increase in the response function.
In the case of compressed lithium, it has been shown that the flat
portions of the corresponding Fermi surface, resulting in Fermi surface
nesting, substantially enhance the electron-phonon interaction \cite{deepa,profeta,sanna}. 

In the present case, the calculated Fermi surfaces of hcp and dhcp
yttrium are shown in Fig. 4. In hcp yttrium, the 3rd band gives rise
to a cylindrical tube along $\Gamma$-A, connected to a pancake-like
structure with hexagonal symmetry on either end, while the fourth
band gives rise to a cylindrical tube along $\Gamma$-A with a bigger
radius. Due to pressure, the Fermi surface of yttrium is substantially
altered, as can be seen in Fig. 4 where we also show the Fermi surface
of dhcp yttrium. The most obvious thing to note is the reduction along
$k_{z}$-direction of the Fermi surface of dhcp yttrium by half. It
is due to the doubling of the unit cell along the \emph{z}-direction
as one goes from hcp to dhcp structure. The cylindrical tube has separated
from the pancake-like surface and, in addition, we find some flat
portions on the M-K-H-L plane with \emph{webbing} \cite{crowe} like
features. 

As shown in Figs. 4 (c)-(d), the change in pressure does not substantially
alter the shape of the Fermi surfaces of either hcp or dhcp phase.
However, to be able to compare the Fermi surfaces of hcp and dhcp
phases at ambient pressure, we have calculated the Fermi surface of
hcp phase after doubling the length of its primitive cell along $c-$direction.
A band-by-band comparison of the two Fermi surfaces of hcp and dhcp
phases at ambient pressure is shown in Fig. 5. The smaller values
of $a$ and $c/a$ in hcp phase compared to dhcp phase are responsible
for bigger Brillouin zones corresponding to the hcp phase. The differences
in the Fermi surfaces of the two phases can then be attributed to
the way the four basis atoms are distributed in the respective unit
cells. 

\begin{figure}
\begin{centering}\subfigure[]{\includegraphics[clip,scale=0.25]{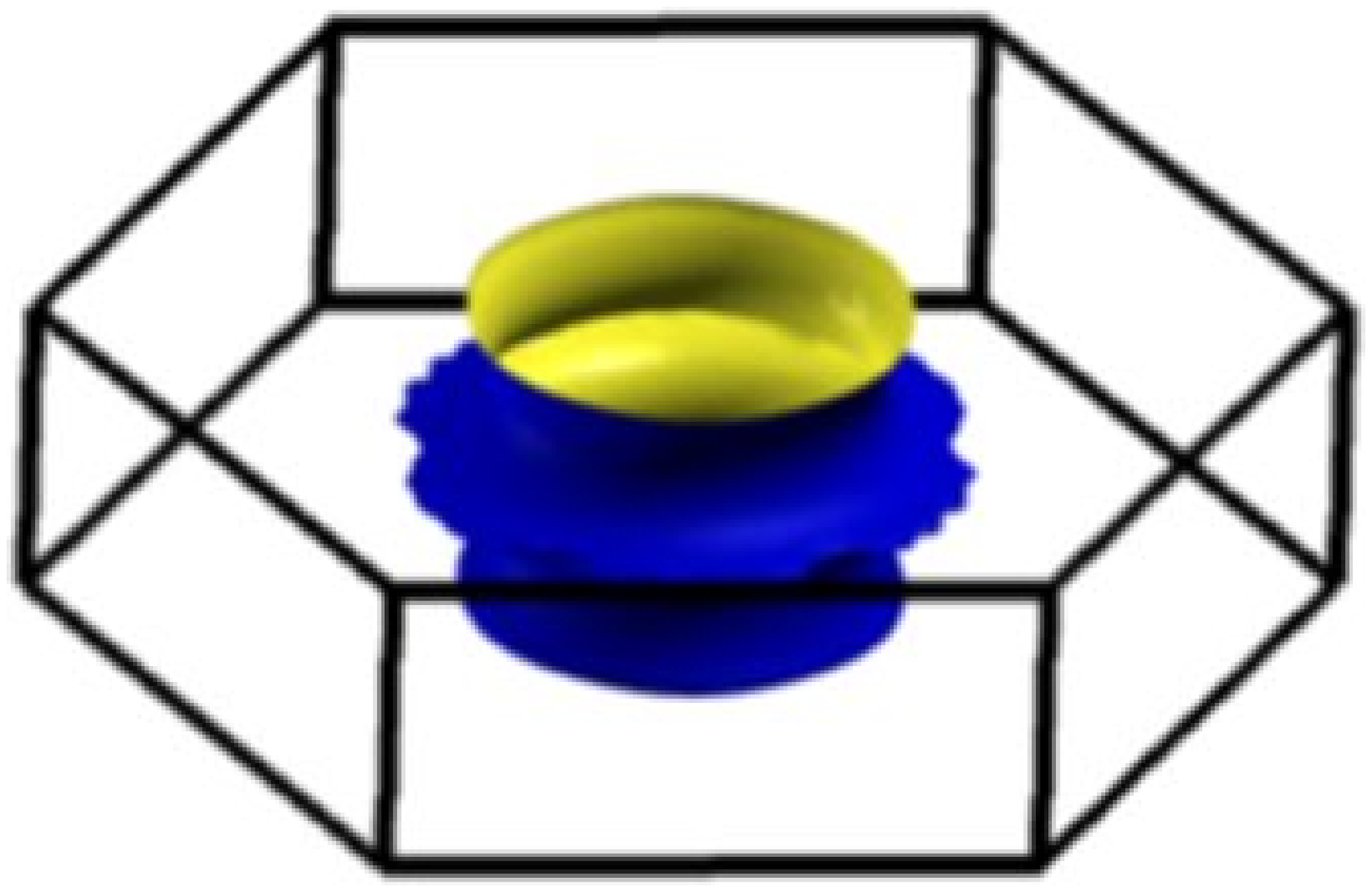}}~\subfigure[]{\includegraphics[scale=0.25]{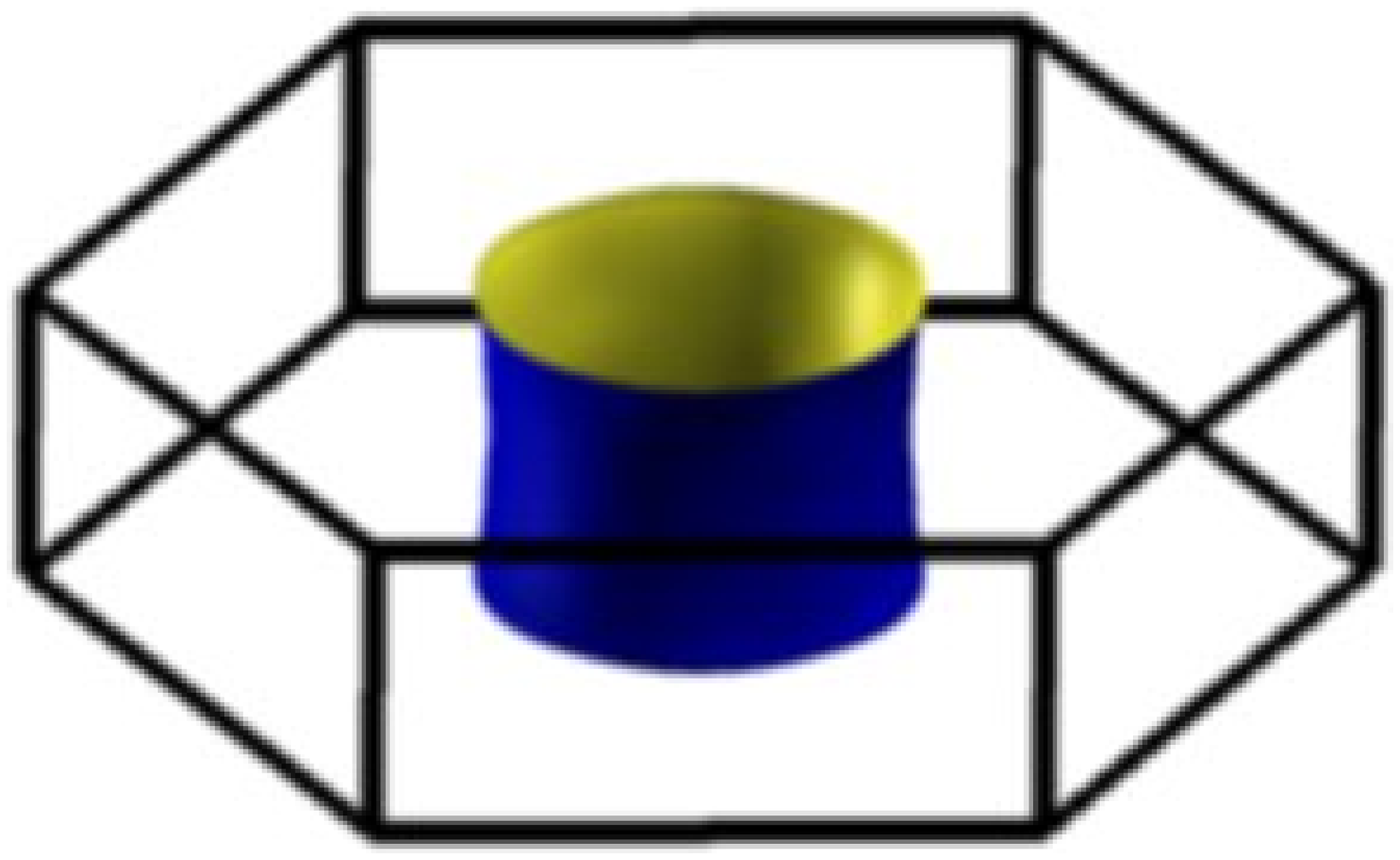}}\par\end{centering}

\begin{centering}\subfigure[]{\includegraphics[clip,scale=0.25]{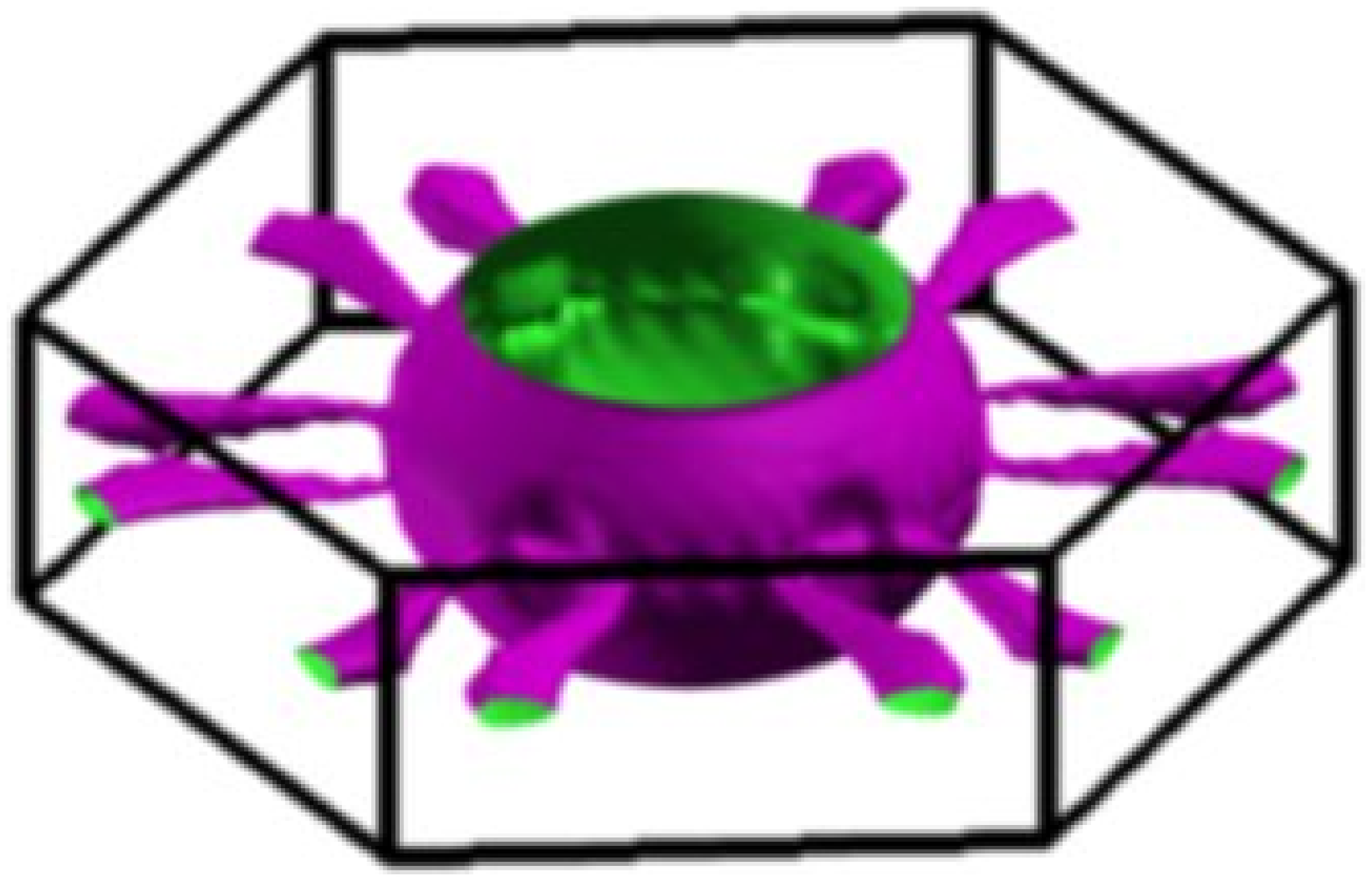}}~\subfigure[]{\includegraphics[scale=0.25]{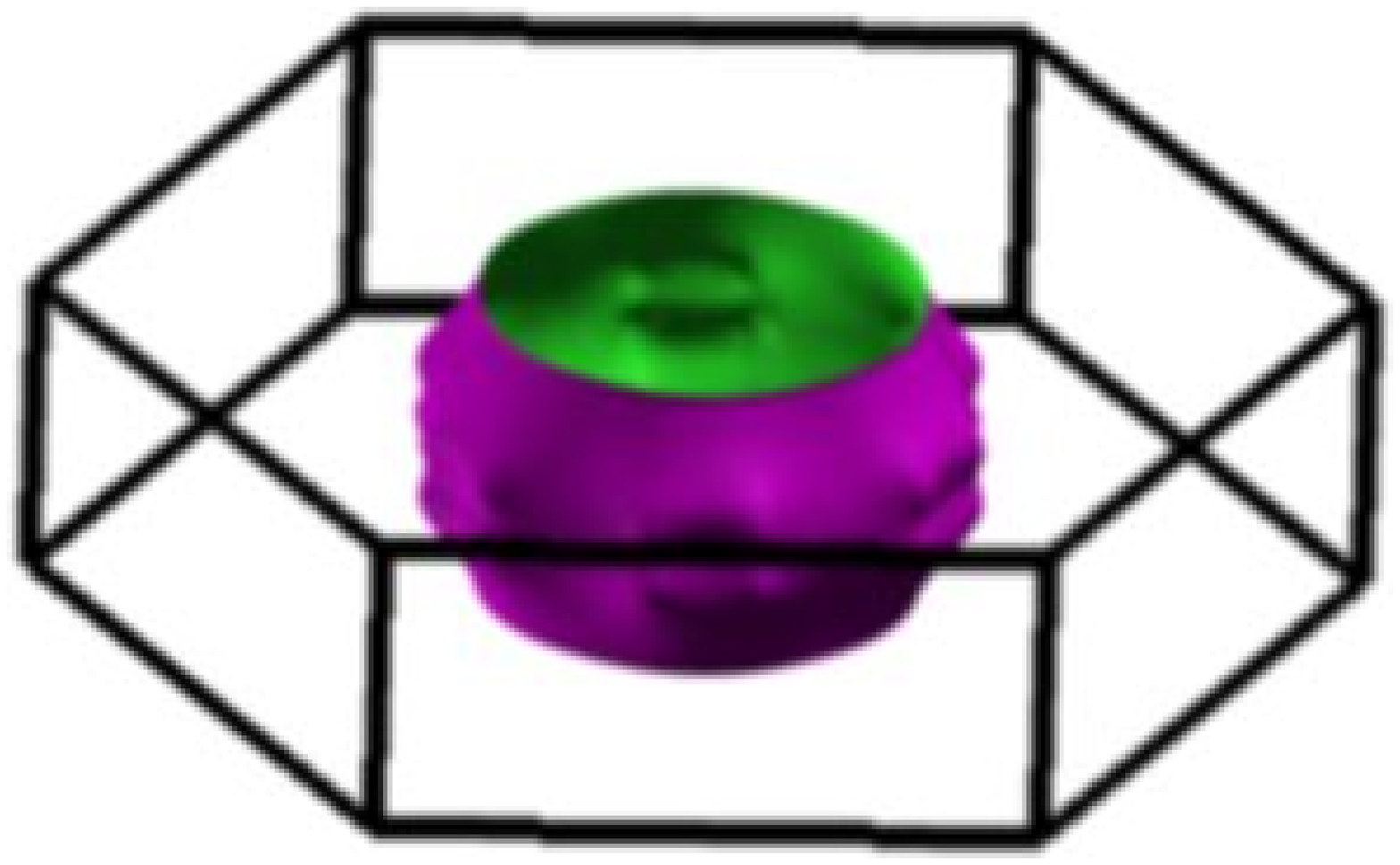}}\par\end{centering}

\begin{centering}\subfigure[]{\includegraphics[clip,scale=0.25]{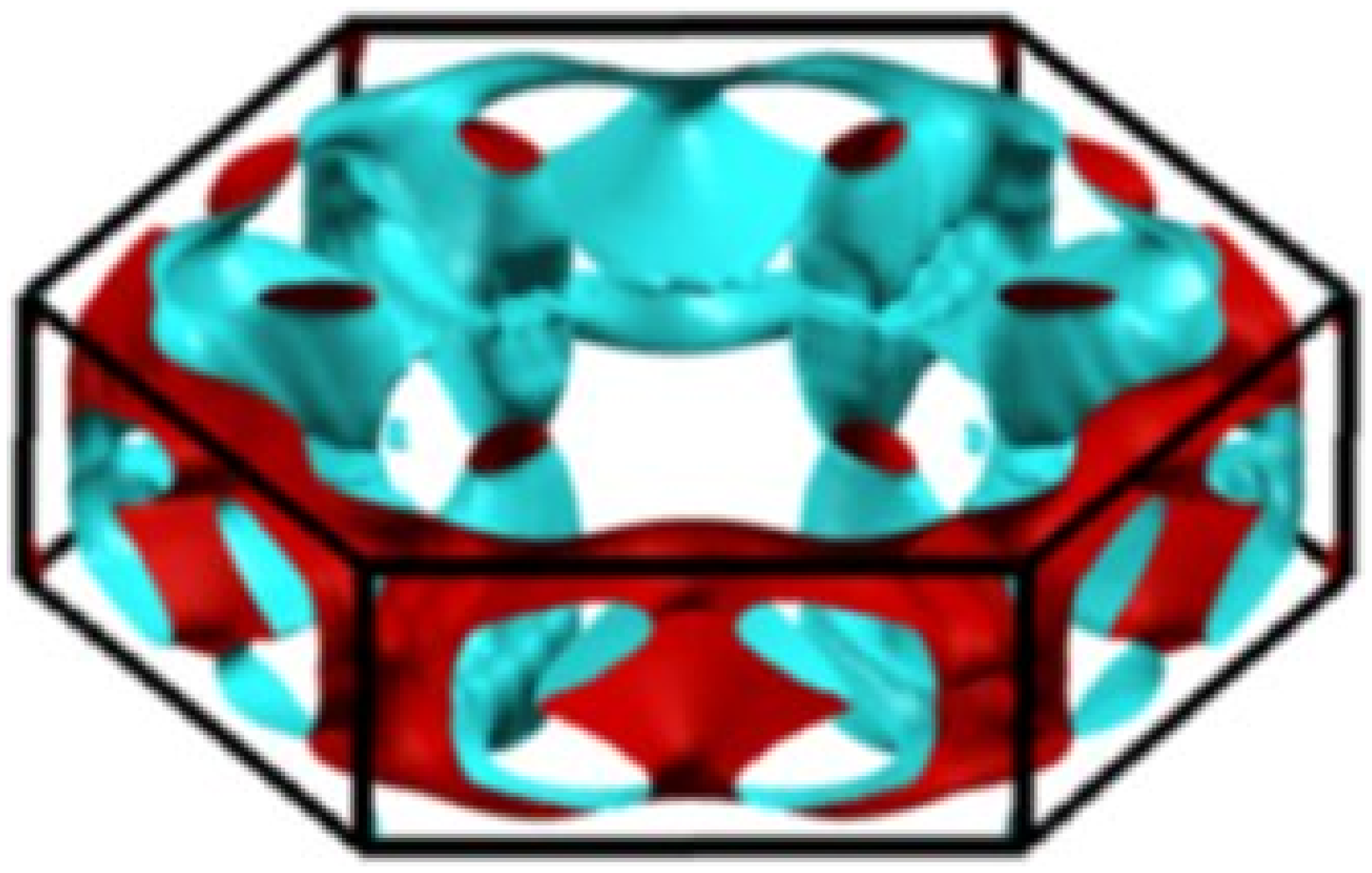}}~\subfigure[]{\includegraphics[scale=0.25]{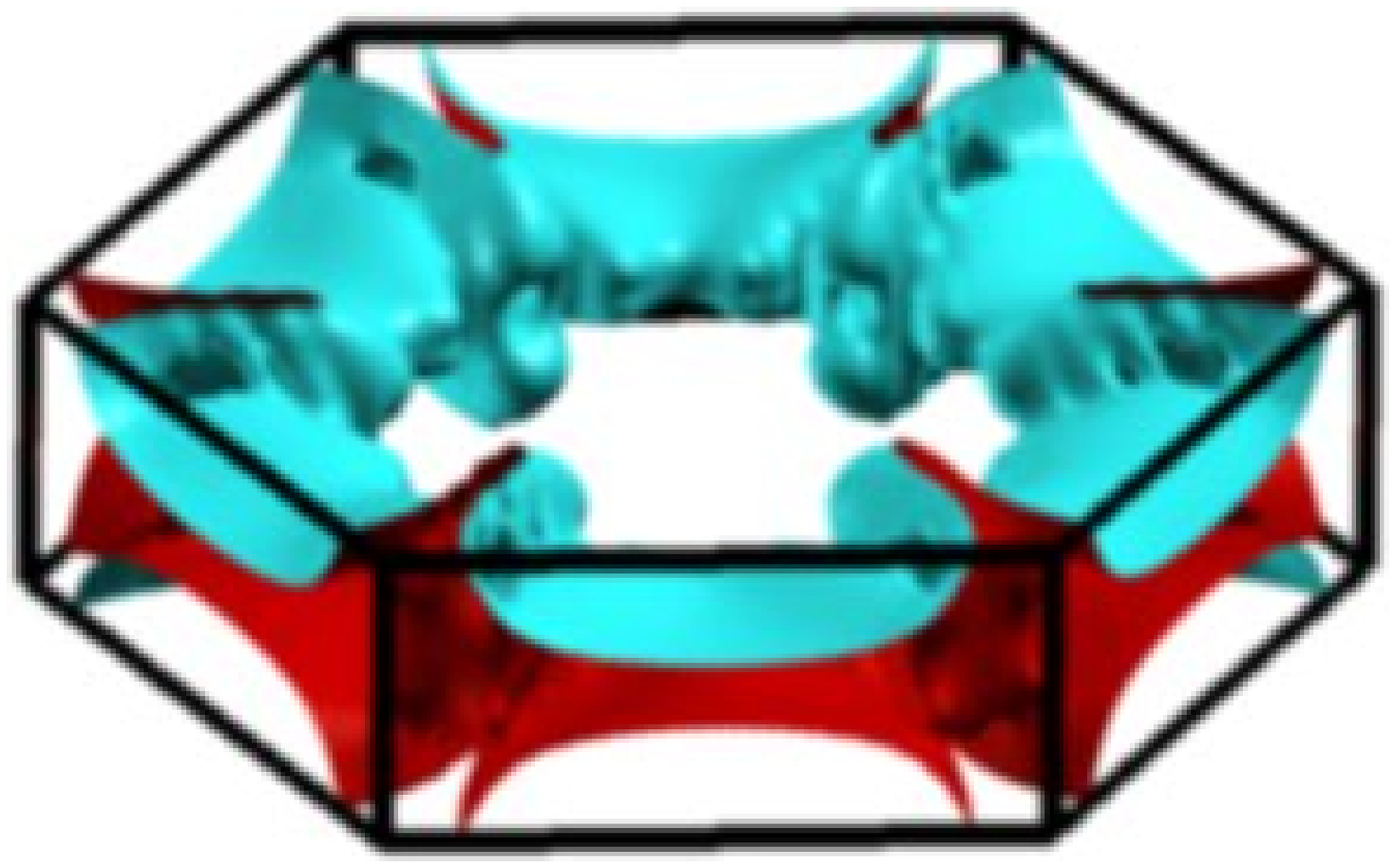}}\par\end{centering}

\begin{centering}\subfigure[]{\includegraphics[scale=0.25]{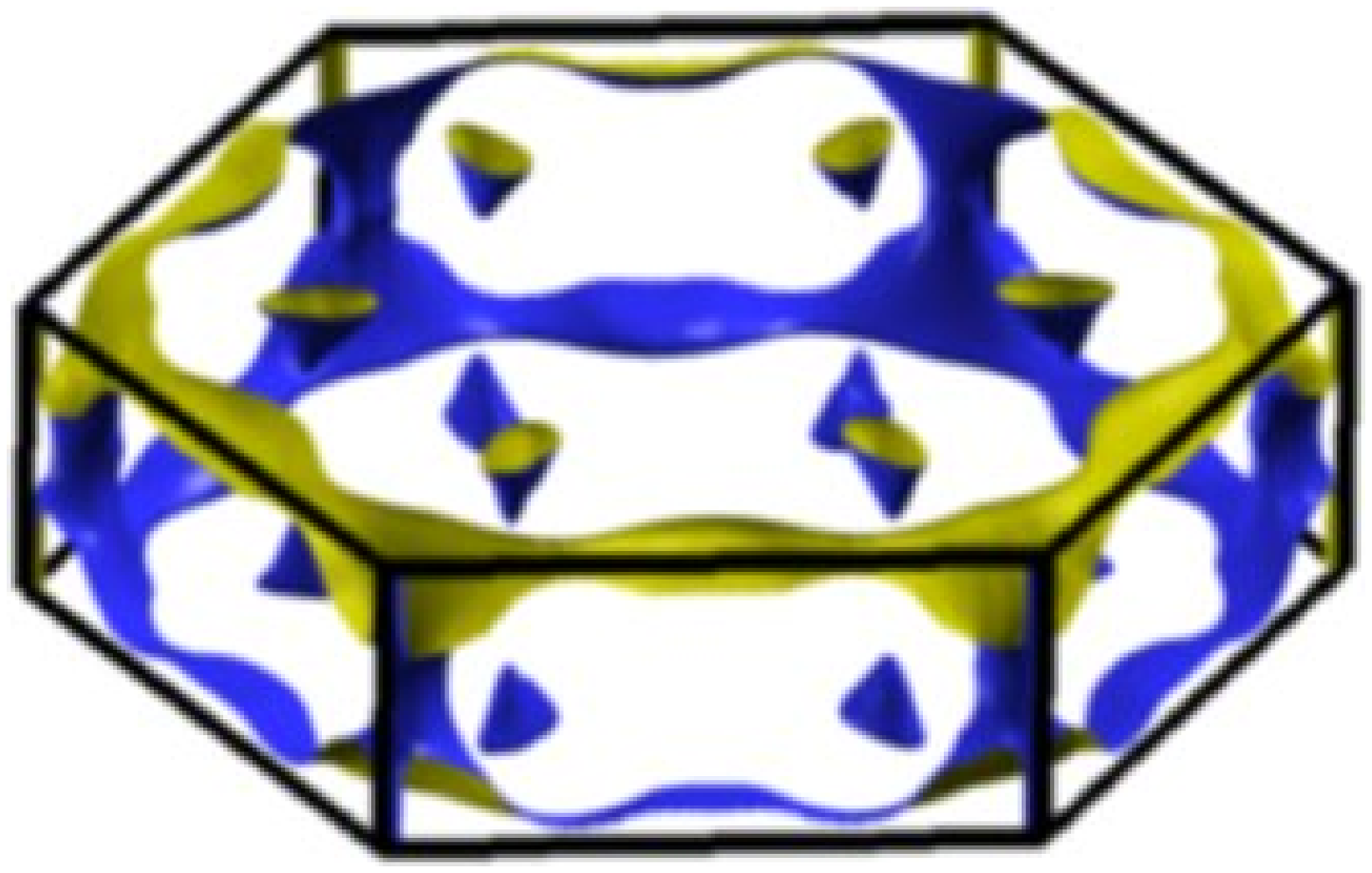}}~\subfigure[]{\includegraphics[scale=0.25]{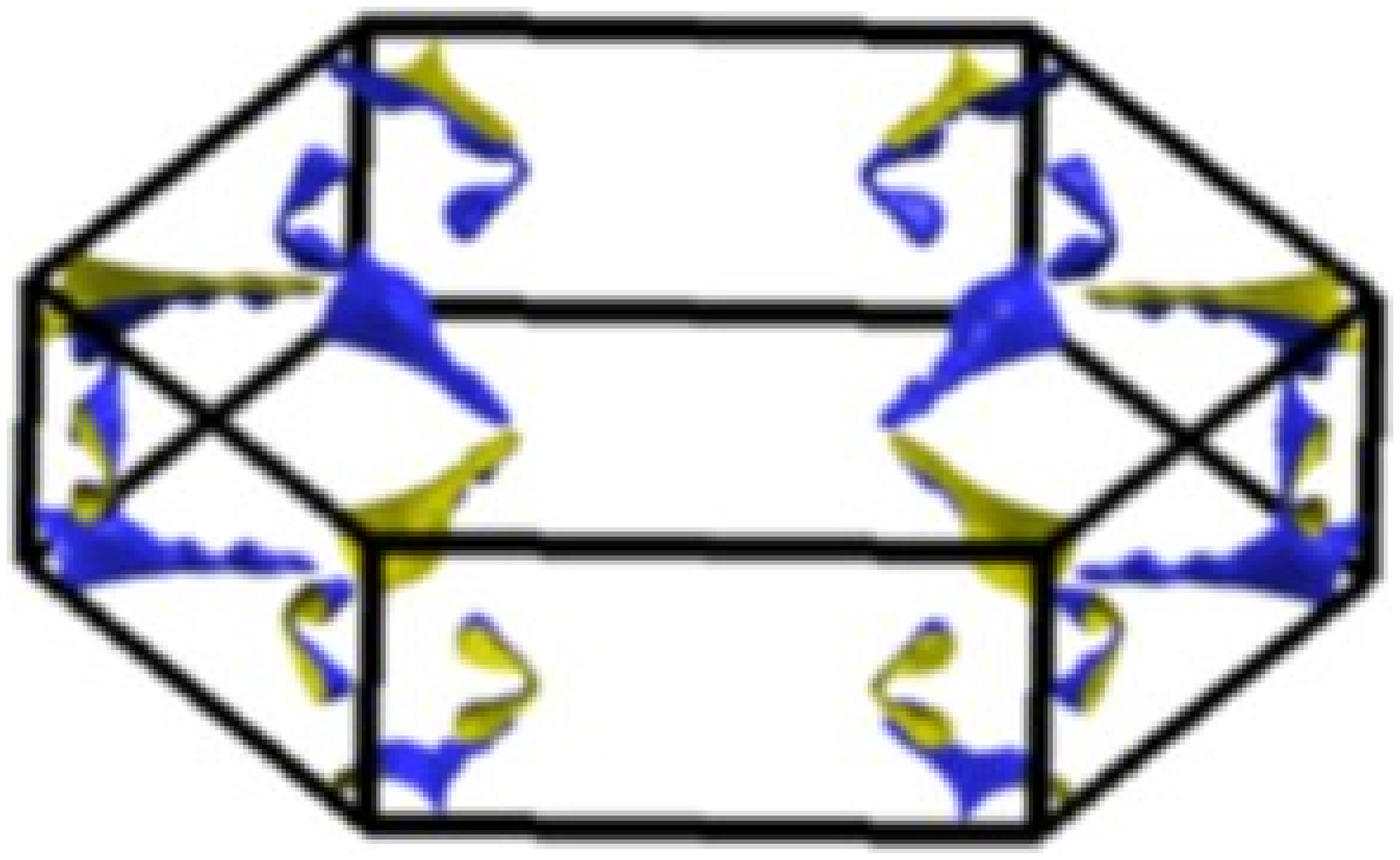}}\par\end{centering}

\caption{The Fermi surface of hcp (left column) and dhcp (right column) yttrium,
both at ambient pressure, corresponding to (a-b) band number 1, (c-d)
band number 2, (e-f) band number 3, and (g-h) band number 4.}
\end{figure}

\subsection{Electron-Phonon Interaction}

The primary objective of the present study is to understand the pressure-induced
increase in the superconducting transition temperature $T_{c}$ of
yttrium. Within the conventional BCS theory of superconductivity,
a change in $T_{c}$ arises due to changes in the phonon spectrum
and/or the electron-phonon spectral function (also known as the Eliashberg
function). Usually, with increasing pressure the lattice hardens resulting
in an increase in the phonon frequencies. At the same time, it is
also possible for some phonon modes to soften and, in some cases,
it may lead to lattice instability. 

\begin{figure}
\begin{centering}\subfigure[]{\includegraphics[clip,scale=0.33]{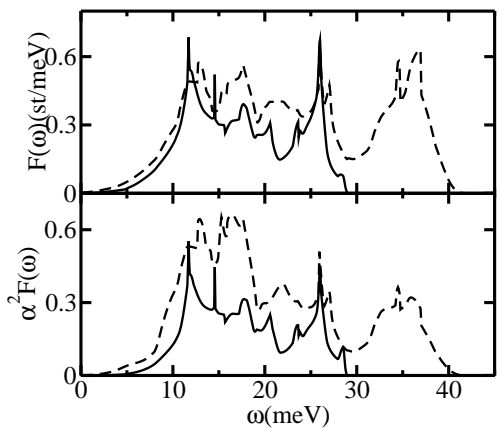}}\par\end{centering}

\begin{centering}\subfigure[]{\includegraphics[clip,scale=0.33]{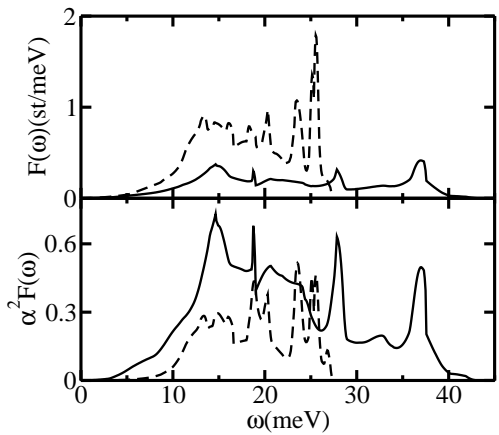}}\par\end{centering}

\caption{The phonon density of states $F(\omega)$ (top panels in (a) and
(b)) and the Eliashberg function $\alpha^{2}F(\omega)$ (bottom panels
in (a) and (b)) of (a) hcp yttrium at ambient pressure (solid line)
and dhcp yttrium at $50$ GPa (dashed line), and (b) hcp yttrium at
$50$ GPa (solid line) and dhcp yttrium at ambient pressure (dashed
line).}
\end{figure}

In Fig. 6, we show the phonon density of states $F(\omega)$ of stable
hcp and dhcp yttrium, calculated using the density-functional theory
as described earlier. The phonon density of states of hcp yttrium,
as given in Fig. 6, is somewhat harder than the results of Heid \emph{et
al} \cite{heid}. The maximum phonon frequency is about $29$ meV
with a large peak centered at $26$ meV. We find that for $\mathbf{q}=\mathbf{0}$,
the maximum frequency has a contribution from the two basis atoms
moving along the \emph{z}-axis in opposite directions. By comparing
the $F(\omega)$ of dhcp yttrium with that of stable hcp phase, we
clearly see the hardening of the lattice as reflected in the increase
of the maximum frequency to $41.8$ meV with a broad peak at around
$36.9$ meV. We find that for $\mathbf{q}=\mathbf{0}$, the maximum
frequency has contribution from the four basis atoms, separated into
two groups of equivalent atoms, both groups moving along the \emph{z}-axis
in opposite directions.

To see the pressure-induced changes in the electron-phonon interaction
in yttrium, we have calculated the Eliashberg function $\alpha^{2}F(\omega)$
of hcp and dhcp yttrium as shown in Fig. 6. A comparison of $\alpha^{2}F(\omega)$
of hcp and dhcp yttrium shows that the electron-phonon coupling is
weaker in hcp yttrium than in dhcp yttrium. In particular, the average
electron-phonon coupling constant $\lambda$ is equal to $0.55$ for
hcp yttrium and $1.24$ for dhcp yttrium. The pressure-induced increase
in $\lambda$ occurs due to enhanced coupling between $12-25$ meV
and due to the coupling of those phonons that arise due to the hardening
of the lattice. Thus, we find that the Eliashberg function $\alpha^{2}F(\omega)$
extends up to $41.8$ meV in dhcp yttrium. The higher frequency couplings
are more effective in raising the superconducting transition temperature
than the lower frequency couplings. 

The role played by the crystal structure in determining $F(\omega)$
and $\alpha^{2}F(\omega)$ can be seen from Fig. 6 (b), where we show
$F(\omega)$ and $\alpha^{2}F(\omega)$ for the unstable hcp and dhcp
yttrium phases. Not surprisingly, now it is the ambient pressure dhcp
phase of yttrium which has a lower maximum frequency of $27.5$ meV
with the high pressure hcp phase having a higher maximum frequency
of $43.6$ meV. Consequently, the electron-phonon coupling is weaker
in the dhcp phase than in the hcp phase, leading to $\lambda=0.47$
for dhcp yttrium and $\lambda=1.2$ for hcp yttrium. 

\begin{figure}
\begin{centering}\subfigure[]{\includegraphics[clip,scale=0.33,angle=270]{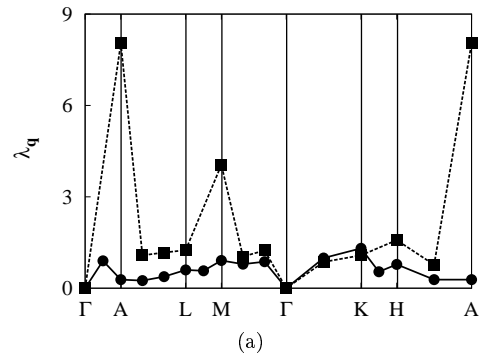}}\par\end{centering}

\begin{centering}\subfigure[]{\includegraphics[clip,scale=0.33,angle=270]{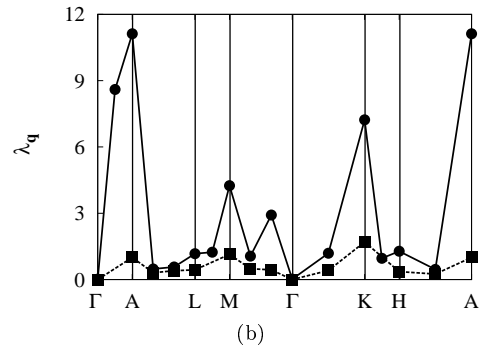}}\par\end{centering}

\caption{The partial $\lambda_{\mathbf{q}}$ along high symmetry directions
in the hexagonal Brillouin zones of (a) hcp yttrium at ambient pressure
(filled circles connected with solid line) and dhcp yttrium at $50$
GPa (filled squares connected with dashed line), and (b) hcp yttrium
at $50$ GPa (filled circles connected with solid line), and dhcp
yttrium at ambient pressure (filled squares connected with dashed
line). The solid and the dotted lines connecting the calculated points
are only a guide to the eye.}
\end{figure}

To gain further insight into the nature of electron-phonon coupling,
we show in Fig. 7 the $\mathbf{q}$-resolved partial $\lambda_{\mathbf{q}}$
along symmetry directions in the corresponding Brillouin zones for
stable and unstable hcp and dhcp phases of yttrium. We find that as
we go from ambient pressure hcp phase to high pressure dhcp phase,
the electron-phonon coupling enhances substantially around symmetry
points A and M, as can be seen from Fig. 7 (a). The contribution to
$\lambda$ from symmetry point A is $\lambda_{\mathbf{q}}=8.1$, which
arises from the lowest eight modes involving motion of all the four
atoms in the $x-y$ plane. Similarly, at the symmetry point M, $\lambda_{\mathbf{q}}=4.1$,
with the lowest mode involving all the four atoms contributing around
$1.6.$ 

A comparison of $\lambda_{\mathbf{q}}$ for unstable dhcp and hcp
phases with that of the stable hcp and dhcp phases shows significant
contributions from symmetry point K, in addition to symmetry points
A and M. The value of $\lambda_{\mathbf{q}}=8.6$ at symmetry point
A arises from the doubly degenerate, bond-stretching mode of the two
basis atoms in the $x-y$ plane, each making a contribution of $3.8$
to $\lambda_{\mathbf{q}}$. 

We like to point out that the integrated values of $\lambda_{\mathbf{q}}$
for the unstable phases may be similar to the stable phases but the
details are very different. We have also seen that a substantial contribution
to $\alpha^{2}F(\omega)$ and consequently to $\lambda$ comes from
a small region in reciprocal space, using a uniform grid to sample
the space requires a dense grid of points. Thus, our calculated values
of $\alpha^{2}F(\omega)$ and consequently, $\lambda$ can be improved
upon by including more \textbf{q}-points in the linear response calculations
but it is unlikely to change the main conclusions of the present work.

\subsection{The Superconducting Transition Temperature}

The possibility of superconductivity in hcp and dhcp yttrium within
the present approach can be checked by solving numerically the isotropic
gap equation \cite{allen2,private1} using the calculated Eliashberg
function $\alpha^{2}F(\omega)$. The results of such a calculation
for hcp and dhcp yttrium for two values of $\mu^{*}$ are shown in
Table II. At ambient pressure, we find hcp yttrium to be superconducting
with $T_{c}=2.8$ K for $\mu^{*}=0.12$ and $T_{c}=0.3$ K for $\mu^{*}=0.20$.
In the high pressure dhcp phase, the $T_{c}$ increases to $19.0$
K for $\mu^{*}=0.12$ and $15.3$ K for $\mu^{*}=0.20$, respectively.
In Table II we have also listed the electron-phonon coupling constant
$\lambda$, and the various averages of phonon frequencies for hcp
and dhcp yttrium at both ambient and $50$ GPa pressure. 

\begin{table}

\caption{The calculated average electron-phonon coupling constant $\lambda$,
the root mean square frequency $\omega_{rms}$ and the superconducting
transition temperature $T_{c}$ obtained with $\mu^{*}=0.2$,  for
yttrium as a function of structure and pressure. The $T_{c}$ values
given in the parentheses correspond to $\mu^{*}=0.12$. }

\begin{centering}\begin{tabular}{|c|c|c|c|c|}
\hline 
yttrium&
pressure (GPa)&
$\lambda$&
$\omega_{rms}$(K) &
$T_{c}$(K)\tabularnewline
\hline
\hline 
hcp&
0&
0.55&
193.5&
0.3 (2.8)\tabularnewline
\hline 
hcp&
50&
1.21&
221.7&
15.9 (19.8)\tabularnewline
\hline 
dhcp&
0&
0.47&
209.0&
$<0.001$(1.4)\tabularnewline
\hline 
dhcp&
50&
1.24&
214.2&
15.3 (19.0)\tabularnewline
\hline
\end{tabular}\par\end{centering}
\end{table}

\section{conclusions}

To understand the pressure-induced changes in electronic structure
and electron-phonon interaction in yttrium, we have studied hexagonal
close-packed \emph{}(hcp) yttrium, stable at ambient pressure and
double hexagonal close-packed \emph{}(dhcp) yttrium, stable up to
around 44 GPa, using density-functional-based methods. Our results
show that as one goes from hcp yttrium to dhcp yttrium, there is (i)
a substantial charge-transfer from $s\rightarrow d$ with extensive
modifications of the \emph{d}-band, (ii) a stiffening of phonon modes,
(iii) an electron-phonon coupling over the entire extended frequency
range with electron-phonon coupling constant $\lambda$ changing from
$0.55$ to $1.24$ and (iv) a change in the superconducting transition
temperature $T_{c}$ from $0.3$ K to $15.3$ K for $\mu^{*}=0.2,$
consistent with experiment.


\begin{thebibliography}{23}
\expandafter\ifx\csname natexlab\endcsname\relax\def\natexlab#1{#1}\fi
\expandafter\ifx\csname bibnamefont\endcsname\relax
  \def\bibnamefont#1{#1}\fi
\expandafter\ifx\csname bibfnamefont\endcsname\relax
  \def\bibfnamefont#1{#1}\fi
\expandafter\ifx\csname citenamefont\endcsname\relax
  \def\citenamefont#1{#1}\fi
\expandafter\ifx\csname url\endcsname\relax
  \def\url#1{\texttt{#1}}\fi
\expandafter\ifx\csname urlprefix\endcsname\relax\def\urlprefix{URL }\fi
\providecommand{\bibinfo}[2]{#2}
\providecommand{\eprint}[2][]{\url{#2}}

\bibitem[{\citenamefont{Probst and Wittig}(1978)}]{probst}
\bibinfo{author}{\bibfnamefont{C.}~\bibnamefont{Probst}} \bibnamefont{and}
  \bibinfo{author}{\bibfnamefont{J.}~\bibnamefont{Wittig}}, in
  \emph{\bibinfo{booktitle}{Handbook on the Physics and Chemistry of Rare
  Earths}}, edited by
  \bibinfo{editor}{\bibfnamefont{J.}~\bibnamefont{K.~A.~Gschneidner}}
  \bibnamefont{and} \bibinfo{editor}{\bibfnamefont{L.}~\bibnamefont{Eyring}}
  (\bibinfo{publisher}{North-Holland}, \bibinfo{address}{Amsterdam},
  \bibinfo{year}{1978}), p. \bibinfo{pages}{749}.

\bibitem[{\citenamefont{Hamlin et~al.}(2006)\citenamefont{Hamlin, Tissen, and
  Schilling}}]{hamlin}
\bibinfo{author}{\bibfnamefont{J.~J.} \bibnamefont{Hamlin}},
  \bibinfo{author}{\bibfnamefont{V.~G.} \bibnamefont{Tissen}},
  \bibnamefont{and} \bibinfo{author}{\bibfnamefont{J.~S.}
  \bibnamefont{Schilling}}, \bibinfo{journal}{Phys. Rev. B}
  \textbf{\bibinfo{volume}{73}}, \bibinfo{pages}{094522}
  (\bibinfo{year}{2006}).

\bibitem[{\citenamefont{Grosshans and Holzapfel}(1992)}]{gro_dhcp}
\bibinfo{author}{\bibfnamefont{W.~A.} \bibnamefont{Grosshans}}
  \bibnamefont{and} \bibinfo{author}{\bibfnamefont{W.~B.}
  \bibnamefont{Holzapfel}}, \bibinfo{journal}{Phys. Rev. B.}
  \textbf{\bibinfo{volume}{45}}, \bibinfo{pages}{5171} (\bibinfo{year}{1992}).

\bibitem[{\citenamefont{Vohra et~al.}(1981)\citenamefont{Vohra, Olijink,
  Grosshans, and Holzapfel}}]{vohra}
\bibinfo{author}{\bibfnamefont{Y.~K.} \bibnamefont{Vohra}},
  \bibinfo{author}{\bibfnamefont{H.}~\bibnamefont{Olijink}},
  \bibinfo{author}{\bibfnamefont{W.}~\bibnamefont{Grosshans}},
  \bibnamefont{and} \bibinfo{author}{\bibfnamefont{W.~P.}
  \bibnamefont{Holzapfel}}, \bibinfo{journal}{Phys. Rev. Lett.}
  \textbf{\bibinfo{volume}{47}}, \bibinfo{pages}{1065} (\bibinfo{year}{1981}).

\bibitem[{\citenamefont{Shimizu et~al.}(2002)\citenamefont{Shimizu, Kimura,
  Takao, and Amaya}}]{shimuzu}
\bibinfo{author}{\bibfnamefont{K.}~\bibnamefont{Shimizu}},
  \bibinfo{author}{\bibfnamefont{H.}~\bibnamefont{Kimura}},
  \bibinfo{author}{\bibfnamefont{D.}~\bibnamefont{Takao}}, \bibnamefont{and}
  \bibinfo{author}{\bibfnamefont{K.}~\bibnamefont{Amaya}},
  \bibinfo{journal}{Nature} \textbf{\bibinfo{volume}{419}},
  \bibinfo{pages}{597} (\bibinfo{year}{2002}).

\bibitem[{\citenamefont{Struzhkin et~al.}(2002)\citenamefont{Struzhkin,
  Eremets, Gan, Mao, and Hemely}}]{stru}
\bibinfo{author}{\bibfnamefont{V.~V.} \bibnamefont{Struzhkin}},
  \bibinfo{author}{\bibfnamefont{M.~I.} \bibnamefont{Eremets}},
  \bibinfo{author}{\bibfnamefont{W.}~\bibnamefont{Gan}},
  \bibinfo{author}{\bibfnamefont{H.-K.} \bibnamefont{Mao}}, \bibnamefont{and}
  \bibinfo{author}{\bibfnamefont{R.~J.} \bibnamefont{Hemely}},
  \bibinfo{journal}{Science} \textbf{\bibinfo{volume}{298}},
  \bibinfo{pages}{1213} (\bibinfo{year}{2002}).

\bibitem[{\citenamefont{Kasinathan et~al.}(2006)\citenamefont{Kasinathan,
  Kunes, Lazicki, Rosner, Yoo, Scalettar, and Pickett}}]{deepa}
\bibinfo{author}{\bibfnamefont{D.}~\bibnamefont{Kasinathan}},
  \bibinfo{author}{\bibfnamefont{J.}~\bibnamefont{Kunes}},
  \bibinfo{author}{\bibfnamefont{A.}~\bibnamefont{Lazicki}},
  \bibinfo{author}{\bibfnamefont{H.}~\bibnamefont{Rosner}},
  \bibinfo{author}{\bibfnamefont{C.~S.} \bibnamefont{Yoo}},
  \bibinfo{author}{\bibfnamefont{R.~T.} \bibnamefont{Scalettar}},
  \bibnamefont{and} \bibinfo{author}{\bibfnamefont{W.~E.}
  \bibnamefont{Pickett}}, \bibinfo{journal}{Phys. Rev. Lett.}
  \textbf{\bibinfo{volume}{96}}, \bibinfo{pages}{047004}
  (\bibinfo{year}{2006}).

\bibitem[{\citenamefont{Profeta et~al.}(2006)\citenamefont{Profeta, Franchini,
  Lathiotakis, Floris, Sanna, Marques, Luders, Massidda, Gross, and
  Continenza}}]{profeta}
\bibinfo{author}{\bibfnamefont{G.}~\bibnamefont{Profeta}},
  \bibinfo{author}{\bibfnamefont{C.}~\bibnamefont{Franchini}},
  \bibinfo{author}{\bibfnamefont{N.~N.} \bibnamefont{Lathiotakis}},
  \bibinfo{author}{\bibfnamefont{A.}~\bibnamefont{Floris}},
  \bibinfo{author}{\bibfnamefont{A.}~\bibnamefont{Sanna}},
  \bibinfo{author}{\bibfnamefont{M.~A.~L.} \bibnamefont{Marques}},
  \bibinfo{author}{\bibfnamefont{M.}~\bibnamefont{Luders}},
  \bibinfo{author}{\bibfnamefont{S.}~\bibnamefont{Massidda}},
  \bibinfo{author}{\bibfnamefont{E.~K.~U.} \bibnamefont{Gross}},
  \bibnamefont{and}
  \bibinfo{author}{\bibfnamefont{A.}~\bibnamefont{Continenza}},
  \bibinfo{journal}{Phys. Rev. Lett.} \textbf{\bibinfo{volume}{96}},
  \bibinfo{pages}{047004} (\bibinfo{year}{2006}).

\bibitem[{\citenamefont{Sanna et~al.}(2006)\citenamefont{Sanna, Franchini,
  Floris, G.Profeta, Lathiotakis, Luders, Marques, Gross, Continenza, and
  Massidda}}]{sanna}
\bibinfo{author}{\bibfnamefont{A.}~\bibnamefont{Sanna}},
  \bibinfo{author}{\bibfnamefont{C.}~\bibnamefont{Franchini}},
  \bibinfo{author}{\bibfnamefont{A.}~\bibnamefont{Floris}},
  \bibinfo{author}{\bibnamefont{G.Profeta}},
  \bibinfo{author}{\bibfnamefont{N.~N.} \bibnamefont{Lathiotakis}},
  \bibinfo{author}{\bibfnamefont{M.}~\bibnamefont{Luders}},
  \bibinfo{author}{\bibfnamefont{M.~A.~L.} \bibnamefont{Marques}},
  \bibinfo{author}{\bibfnamefont{E.~K.~U.} \bibnamefont{Gross}},
  \bibinfo{author}{\bibfnamefont{A.}~\bibnamefont{Continenza}},
  \bibnamefont{and} \bibinfo{author}{\bibfnamefont{S.}~\bibnamefont{Massidda}},
  \bibinfo{journal}{Phys. Rev. B} \textbf{\bibinfo{volume}{73}},
  \bibinfo{pages}{144512} (\bibinfo{year}{2006}).

\bibitem[{\citenamefont{Melsen et~al.}(1993)\citenamefont{Melsen, Wills,
  Johansson, and Eriksson}}]{melsen}
\bibinfo{author}{\bibfnamefont{J.}~\bibnamefont{Melsen}},
  \bibinfo{author}{\bibfnamefont{J.~M.} \bibnamefont{Wills}},
  \bibinfo{author}{\bibfnamefont{B.}~\bibnamefont{Johansson}},
  \bibnamefont{and} \bibinfo{author}{\bibfnamefont{O.}~\bibnamefont{Eriksson}},
  \bibinfo{journal}{Phys. Rev. B.} \textbf{\bibinfo{volume}{48}},
  \bibinfo{pages}{15574} (\bibinfo{year}{1993}).

\bibitem[{\citenamefont{Wittig}(1970)}]{wittig}
\bibinfo{author}{\bibfnamefont{J.}~\bibnamefont{Wittig}},
  \bibinfo{journal}{Phys. Rev. Lett.} \textbf{\bibinfo{volume}{24}},
  \bibinfo{pages}{812} (\bibinfo{year}{1970}).

\bibitem[{\citenamefont{Savrasov}(1996)}]{savrasov1}
\bibinfo{author}{\bibfnamefont{S.~Y.} \bibnamefont{Savrasov}},
  \bibinfo{journal}{Phys. Rev. B} \textbf{\bibinfo{volume}{54}},
  \bibinfo{pages}{16470} (\bibinfo{year}{1996}).

\bibitem[{\citenamefont{Savrasov and Savrasov}(1996)}]{savrasov2}
\bibinfo{author}{\bibfnamefont{S.~Y.} \bibnamefont{Savrasov}} \bibnamefont{and}
  \bibinfo{author}{\bibfnamefont{D.~Y.} \bibnamefont{Savrasov}},
  \bibinfo{journal}{Phys. Rev. B} \textbf{\bibinfo{volume}{54}},
  \bibinfo{pages}{16487} (\bibinfo{year}{1996}).

\bibitem[{pws()}]{pwscf}
\emph{\bibinfo{title}{plane-wave self-consistent-field code:}},
  \eprint{http://wwww.pwscf.org/}.

\bibitem[{\citenamefont{Allen and Dynes}(1975)}]{allen1}
\bibinfo{author}{\bibfnamefont{P.~B.} \bibnamefont{Allen}} \bibnamefont{and}
  \bibinfo{author}{\bibfnamefont{R.~C.} \bibnamefont{Dynes}},
  \bibinfo{journal}{Phys.\ Rev. B} \textbf{\bibinfo{volume}{12}},
  \bibinfo{pages}{905} (\bibinfo{year}{1975}).

\bibitem[{\citenamefont{Allen and Mitrovic}(1982)}]{allen2}
\bibinfo{author}{\bibfnamefont{P.}~\bibnamefont{Allen}} \bibnamefont{and}
  \bibinfo{author}{\bibfnamefont{B.}~\bibnamefont{Mitrovic}}, in
  \emph{\bibinfo{booktitle}{Advances in Solid State Physics}}, edited by
  \bibinfo{editor}{\bibfnamefont{H.}~\bibnamefont{Ehrenreich}},
  \bibinfo{editor}{\bibfnamefont{F.}~\bibnamefont{Seitz}}, \bibnamefont{and}
  \bibinfo{editor}{\bibfnamefont{E.}~\bibnamefont{Turnbull}}
  (\bibinfo{publisher}{Academic Press}, \bibinfo{address}{New York},
  \bibinfo{year}{1982}), \bibinfo{number}{37}, p.~\bibinfo{pages}{1}.

\bibitem[{\citenamefont{Allen}()}]{private1}
\bibinfo{author}{\bibfnamefont{P.~B.} \bibnamefont{Allen}}, \eprint{private
  communication}.

\bibitem[{\citenamefont{Yin et~al.}()\citenamefont{Yin, Savrasov, and
  Pickett}}]{yin}
\bibinfo{author}{\bibfnamefont{Z.~P.} \bibnamefont{Yin}},
  \bibinfo{author}{\bibfnamefont{S.~Y.} \bibnamefont{Savrasov}},
  \bibnamefont{and} \bibinfo{author}{\bibfnamefont{W.~E.}
  \bibnamefont{Pickett}}, \eprint{cond-mat/0606538}.

\bibitem[{\citenamefont{Perdew and Wang}(1992)}]{perdew1}
\bibinfo{author}{\bibfnamefont{J.~P.} \bibnamefont{Perdew}} \bibnamefont{and}
  \bibinfo{author}{\bibfnamefont{Y.}~\bibnamefont{Wang}},
  \bibinfo{journal}{Phys.\ Rev. B} \textbf{\bibinfo{volume}{45}},
  \bibinfo{pages}{13244} (\bibinfo{year}{1992}).

\bibitem[{xcr()}]{xcrys}
\emph{\bibinfo{title}{xcrysden code:}}, \eprint{http://wwww.xcrysden.org/}.

\bibitem[{\citenamefont{Birch}(1947)}]{birch}
\bibinfo{author}{\bibfnamefont{F.}~\bibnamefont{Birch}},
  \bibinfo{journal}{Phys. Rev.} \textbf{\bibinfo{volume}{71}},
  \bibinfo{pages}{809} (\bibinfo{year}{1947}).

\bibitem[{\citenamefont{Crowe et~al.}(2004)\citenamefont{Crowe, Dugdale, Major,
  Alam, Duffy, and Palmer}}]{crowe}
\bibinfo{author}{\bibfnamefont{S.~J.} \bibnamefont{Crowe}},
  \bibinfo{author}{\bibfnamefont{S.~B.} \bibnamefont{Dugdale}},
  \bibinfo{author}{\bibfnamefont{Z.}~\bibnamefont{Major}},
  \bibinfo{author}{\bibfnamefont{M.~A.} \bibnamefont{Alam}},
  \bibinfo{author}{\bibfnamefont{J.~A.} \bibnamefont{Duffy}}, \bibnamefont{and}
  \bibinfo{author}{\bibfnamefont{S.~B.} \bibnamefont{Palmer}},
  \bibinfo{journal}{Europhys. Lett.} \textbf{\bibinfo{volume}{65}},
  \bibinfo{pages}{235} (\bibinfo{year}{2004}).

\bibitem[{\citenamefont{Heid and Bohnen}(1999)}]{heid}
\bibinfo{author}{\bibfnamefont{R.}~\bibnamefont{Heid}} \bibnamefont{and}
  \bibinfo{author}{\bibfnamefont{K.-P.} \bibnamefont{Bohnen}},
  \bibinfo{journal}{Phys. Rev. B} \textbf{\bibinfo{volume}{60}},
  \bibinfo{pages}{R3709} (\bibinfo{year}{1999}).

\end{thebibliography}
\end{document}